\newif\ifnotes
\notestrue 

\documentclass[10pt,journal,compsoc]{IEEEtran}

\usepackage{algpseudocode}

\usepackage[normalem]{ulem}
\usepackage{setspace}

\usepackage[usenames,dvipsnames,table]{xcolor}

\ifnotes
  \usepackage[nomargin,inline,draft]{fixme}
\else
  \usepackage[nomargin,inline,final]{fixme}
\fi

\usepackage[anythingbreaks]{breakurl}
\usepackage[
  sorting=none,
  style=numeric-comp,
  backend=biber,
  isbn=false,
  url=true,
  doi=false,
  url=false,
  mincrossrefs=100,
  maxnames=12,
  firstinits=true
]{biblatex}

\usepackage[utf8]{inputenc}
\usepackage[T1]{fontenc}
\usepackage{microtype}
\usepackage{pstricks}
\usepackage{graphicx}
\usepackage{amsmath}
\usepackage{amssymb}
\usepackage{xspace}
\usepackage{listings}
\usepackage{multirow}

\usepackage{amsthm}

\usepackage{mathtools}
\DeclarePairedDelimiter{\ceil}{\lceil}{\rceil}
\DeclarePairedDelimiter{\floor}{\lfloor}{\rfloor}

\usepackage{relsize}

\usepackage{caption}
\usepackage{subcaption}
\usepackage{mdframed}

\newcommand\omnetpp{Omnet\nolinebreak[4]\hspace{-.05em}\raisebox{.4ex}{\relsize{-3}{\textbf{++}}}~}
\newcommand\cpp{C\nolinebreak[4]\hspace{-.05em}\raisebox{.4ex}{\relsize{-3}{\textbf{++}}}~}

\AtEveryBibitem{\clearname{editor}}

\definecolor{vrpink}{RGB}{255,0,127}
\definecolor{vrblue}{RGB}{30,144,255}
\definecolor{vrolive}{RGB}{85,107,47}
\definecolor{vrroyalblue}{RGB}{65,105,225}
\definecolor{brgreen}{RGB}{100,200,70}
\definecolor{ivsalmon}{RGB}{255,160,122}
\definecolor{vrlpink}{RGB}{255,192,203}
\definecolor{mvcol}{RGB}{5,150,25}

\usepackage[obeyFinal]{todonotes}
\FXRegisterAuthor{vv}{avr}{\color{vrpink}[vince]}
\FXRegisterAuthor{rvv}{arvr}{\color{red}[vince]}

\newcommand{\newtext}[1]{#1}

\fxusetheme{color}
\fxuseenvlayout{color}

\usepackage{graphicx}
\graphicspath{{./figures/}}
\DeclareGraphicsExtensions{.pdf,.jpeg,.png,.jpg}

\usepackage{algorithm}
\usepackage{algpseudocode}
\algblockdefx{Event}{EndEvent}[1]%
{\textbf{upon event }#1 \textbf{do }}%
{\textbf{}}

\algblockdefx{Function}{EndFunction}[1]%
{\textbf{Function }#1 }%
{\textbf{}}

\newcommand{\comalgo}[1]{\textit{\textcolor{Maroon}{#1}}}

\usepackage[
  colorlinks=true,
  linkcolor=purple,
  citecolor=purple,
  urlcolor=black,
  pdfauthor={},
  pdftitle={},
  pdfsubject={},
  pdfkeywords={},
  bookmarks=false,
]{hyperref}

\hypersetup{colorlinks=true}

\makeatletter
\renewenvironment{proof}[1][\proofname]{\par
  \pushQED{\qed}%
  \normalfont
  \topsep0pt \partopsep0pt 
  \trivlist
  \item[\hskip\labelsep
        \itshape
    #1\@addpunct{.}]\ignorespaces
}{%
  \endtrivlist\@endpefalse
}
\makeatother

\usepackage{pgfplotstable}
\pgfplotsset{compat=1.16}
\usepgfplotslibrary{statistics}

\pgfplotstableread{%
Mod low ql med qh high
MBD.1 -98.8 -98.1 -97.8 -97.6 -96.8
MBD.2 -17.1 -4.8 -0.8 3.4 15.7
MBD.3 -15.7 -3.2 0.9 5.1 17.6
MBD.4 -15.7 -4.6 -0.3 2.8 13.9
MBD.5 -18.0 -4.7 -0.3 4.2 17.6
MBD.6 -14.8 -4.3 -0.3 2.7 13.2
MBD.7 -50.4 -33.6 -28.5 -22.4 -5.6
MBD.8 -32.1 -14.7 -8.8 -3.1 14.3
MBD.9 -24.6 -8.6 -4.3 2.1 18.2
MBD.10 -16.5 -4.8 -0.3 3.0 14.7
MBD.11 -68.6 -32.2 -25.0 -7.9 28.6
MBD.12 -32.7 -0.7 7.0 20.6 52.6
}\datatablenetsync

\pgfplotstableread{%
Mod low ql med qh high
MBD.1 -216.8 -97.4 -94.1 -17.8 101.6
MBD.2 -21.5 -5.0 -0.8 6.0 22.5
MBD.3 -16.8 -4.6 -0.3 3.5 15.6
MBD.4 -19.3 -5.5 -0.3 3.7 17.5
MBD.5 -18.7 -4.3 0.2 5.3 19.7
MBD.6 -18.2 -5.1 -0.5 3.6 16.6
MBD.7 -64.6 -25.2 -4.4 1.1 40.6
MBD.8 -28.1 -10.4 -2.3 1.4 19.1
MBD.9 -17.6 -4.9 -0.3 3.6 16.4
MBD.10 -18.3 -3.6 0.3 6.2 20.9
MBD.11 -51.9 -5.7 4.2 25.1 71.3
MBD.12 -40.0 -1.2 7.6 24.7 63.6
}\datatablelatsync

\pgfplotstableread{%
Mod low ql med qh high
MBD.1 -100.2 -97.5 -96.8 -95.7 -93.0
MBD.2 -29.6 -8.6 -1.9 5.4 26.4
MBD.3 -36.6 -10.7 0.5 6.6 32.6
MBD.4 -32.0 -7.8 0.4 8.3 32.4
MBD.5 -19.8 -3.5 1.2 7.4 23.8
MBD.6 -23.0 -6.5 0.1 4.5 21.0
MBD.7 -55.2 -29.6 -24.3 -12.5 13.2
MBD.8 -33.6 -13.2 -6.7 0.4 20.8
MBD.9 -33.5 -9.2 -4.3 7.0 31.3
MBD.10 -33.9 -10.2 -0.7 5.6 29.3
MBD.11 -85.8 -34.0 -17.9 0.5 52.2
MBD.12 -35.9 -4.1 5.1 17.1 48.9
}\datatablenetasync

\pgfplotstableread{%
Mod low ql med qh high
MBD.1 -174.8 -93.5 -83.2 -39.3 42.0
MBD.2 -29.2 -8.6 -2.1 5.1 25.6
MBD.3 -25.9 -6.4 1.2 6.6 26.1
MBD.4 -23.8 -5.5 -0.3 6.7 25.0
MBD.5 -27.4 -6.2 0.6 7.9 29.1
MBD.6 -21.9 -5.1 -0.6 6.1 22.9
MBD.7 -52.8 -24.6 -17.3 -5.8 22.4
MBD.8 -32.6 -13.1 -4.4 -0.1 19.4
MBD.9 -29.4 -8.7 -2.6 5.1 25.8
MBD.10 -22.7 -5.3 -0.1 6.3 23.7
MBD.11 -42.8 -10.9 0.2 10.4 42.4
MBD.12 -37.4 -2.2 6.5 21.3 56.6
}\datatablelatasync

\usepackage{rotating}

\begin{document}

    \sloppy

    \title{Practical Byzantine Reliable Broadcast on \\Partially Connected Networks (Extended version)}
    
\author{
\IEEEauthorblockN{
   	Silvia Bonomi\IEEEauthorrefmark{1},
   	J\'er\'emie Decouchant\IEEEauthorrefmark{2}\thanks{\IEEEauthorrefmark{2} Corresponding author.},
   	Giovanni Farina\IEEEauthorrefmark{1}, 
   	Vincent Rahli\IEEEauthorrefmark{3},
S\'{e}bastien Tixeuil\IEEEauthorrefmark{4}}
   	
\IEEEauthorblockA{
\IEEEauthorrefmark{1}Sapienza Università di Roma,
\IEEEauthorrefmark{2}Delft University of Technology, \\ 
\IEEEauthorrefmark{3}University of Birmingham, 
\IEEEauthorrefmark{4}Sorbonne Universit\'{e}, CNRS, LIP6, Institut Universitaire de France \\
\url{bonomi@diag.uniroma1.it}, \url{j.decouchant@tudelft.nl}, \url{gfarina@diag.uniroma1.it}, \\ \url{vincent.rahli@gmail.com},  \url{sebastien.tixeuil@lip6.fr}
}
}
  
\maketitle

\begin{abstract}
    We consider the Byzantine reliable broadcast problem on authenticated and partially connected networks. 
    The state-of-the-art method to solve this problem consists in combining two algorithms from the literature. 
    Handling asynchrony and faulty senders is typically done thanks to Gabriel Bracha's authenticated double-echo broadcast protocol, which assumes an asynchronous fully connected network. Danny Dolev's algorithm can then be used to provide reliable communications between processes in the global fault model, where up to $f$ processes among $N$ can be faulty in a communication network that is at least $2f{+}1$-connected. 
    Following recent works that showed how Dolev's protocol can be made more practical thanks to several optimizations, we show that the state-of-the-art methods to solve our problem can be \newtext{further} optimized thanks to layer-specific and cross-layer optimizations.  
    \newtext{Our performance evaluation based on real \cpp implementation and actual deployment} show that these optimizations can be efficiently combined to decrease the total amount of information transmitted or the protocol's latency (e.g., respectively, -98\% and -97\% with a 1024~B payload, $N{=}50$ and $f{=}5$) compared to the state-of-the-art combination of Bracha's and Dolev's protocols. 
\end{abstract}

\section{Introduction}
\label{sec:introduction}

At their very core, distributed systems consist of autonomous computing entities (or processes) that \emph{communicate} to globally solve non-trivial tasks. Over the years, distributed systems became ubiquitous and increased the possibility that some of the processes fail in some unpredictable manner. The most general fault model is the Byzantine one, that allows a process to simply exhibit arbitrary behavior. Byzantine processes are opposed to correct processes, that trustfully follow their prescribed algorithm. In this context, communicating reliably may become difficult, especially when processes have to rely on other (possibly faulty) processes to convey the information they want to send (that is, the network is partially connected). 

Two useful global communication abstractions have been defined in this context. First, the \emph{reliable communication} (RC) abstraction requires that: \emph{(i)} if a correct process broadcasts a message, then all correct processes also deliver this message, and \emph{(ii)} if a message is delivered, then it was indeed sent by its claimed source. Reliable communication is also referred to as \emph{reliable broadcast with honest dealer}, outlining that
the broadcast is achieved in case of a correct source.
Second, the \emph{reliable broadcast} abstraction considers the additional case where the sender of a message may be arbitrarily faulty. In this case, all correct processes deliver the same message or none do. Reliable broadcast, often abbreviated as BRB (Byzantine Reliable Broadcast), guarantees stronger properties than reliable communication, yet both abstractions can be implemented in a fully asynchronous network. 
Recently, protocols that rely on Byzantine reliable broadcast have been used to implement Blockchain consensus~\cite{mostefaoui2015intrusion,yu2019repucoin,decouchant2022damysus} and decentralized payments~\cite{DBLP:conf/dsn/CollinsGKKMPPST20, auvolat2020money, cohen2021tame}.

The most well known asynchronous BRB protocol is probably Gabriel Bracha's algorithm, which is sometimes named double echo authenticated broadcast~\cite{Bracha:iandc:1987}. This protocol assumes a fully connected network of $N$ processes (among which at most $f {<} N/3$ are Byzantine) and authenticated network links. 

Questioning whether it is possible to remove the full connectivity assumption in an asynchronous setting leads to Dolev's RC algorithm~\cite{dolev1981unanimity}, as long as the network is "sufficiently well connected" (a necessary and sufficient condition for RC in arbitrary networks is that the connectivity of the network is at least $2f{+}1$, where $f$ is the number of Byzantine processes). 
However, Dolev's algorithm does not solve BRB, as it does not prevent a Byzantine faulty process from causing only a subset of the correct processes to deliver the same message. 
Nevertheless, Bracha's and Dolev's protocols can be combined to solve BRB in sufficiently well-connected synchronous or asynchronous network topologies~\cite{wang2020asynchronous}.

In this paper, we start from this state-of-the-art solution to the BRB problem in partially connected networks, and make the following contributions: \\
(i) We describe how state-of-the-art optimizations of Dolev's RC algorithm~\cite{bonomi2019multi} can be extended to the BRB combination of Bracha's and Dolev's algorithms. \\
(ii) We present a total of 12 novel modifications that can be applied to the BRB combination of Bracha's and Dolev's algorithms, some of them being cross-layer, demonstrating by experience the practicality of our approach. \\
(iii) We evaluate the impact of each modification on the protocol latency and throughput using \newtext{real deployments} in a large variety of settings. \\
(iv) We detail how these modifications should be combined depending on the network asynchrony and connectivity, and on the payload size to optimize latency and/or throughput, helping future deployments of our solution for specific purposes. 

The remainder of this paper is organized as follows.
Sec.~\ref{sec:relatedwork} discusses the related works.
Sec.~\ref{sec:sysmodel} describes the system model and the BRB problem.
Sec.~\ref{sec:background} provides some background on BRB algorithms on asynchronous partially connected networks.
Sec.~\ref{sec:brachadolev} explains how we modify the interface of the Bracha-Dolev protocol to include state-of-the-art improvements of Dolev's protocol, and to add functionalities.
Sec.~\ref{sec:latency} 
present our modifications MBD.1--12 of the Bracha-Dolev protocol. 
Sec.~\ref{sec:perf_eval} details our performance evaluation.
Finally, Sec.~\ref{sec:conclusion} concludes the paper.

\section{Related Works} \label{sec:relatedwork}

The first protocol to consider the Byzantine Reliable Broadcast (BRB) problem assumed authenticated links\footnote{Let us recall that authenticated links guarantee that the identity of the sender cannot be forged and can also be implemented without cryptography~\cite{zeng2010non}.}.

Bracha and Toueg formalized the Reliable Broadcast problem~\cite{bracha1984asynchronous, Bracha+Toueg:jacm:1985} and then Bracha described the first BRB protocol for asynchronous and fully connected reliable networks (i.e., networks where each process is able to communicate with any other in the system and where messages cannot be lost) and the proposed solution is able to tolerate $f$ Byzantine nodes (where $f < N/3$ and $N$ is the number of processes in the system)~\cite{Bracha:iandc:1987}. 
This protocol is characterized by three different all-to-all communication phases (namely, \emph{send}, \emph{echo} and \emph{ready}) and processes progress in the algorithm as soon as they have heard from a quorum of nodes in a given phase.

This protocol assumes a fully connected communication network and thus its applicability in general networks (like the ones considered in this paper) is not directly possible. 

Concerning generic networks where full connectivity cannot be assumed, Dolev~\cite{dolev1981unanimity} showed that correct processes can reliably communicate in presence of $f$ Byzantine nodes if, and only if, the network is $(2f{+}1)$-connected~\cite{diestel2017}. 
In particular, Dolev's algorithm allows a process $p_i$ to deliver a
message when it receives it through at least $f{+}1$ disjoint paths (which is made possible when it
flows through at least $2f{+}1$ disjoint paths). In order to do that, it
requires to solve a maximum disjoint paths problem. Also in this case,
the solution assumes authenticated and reliable point-to-point
communication links.
Despite its theoretical correctness, Dolev's solution is not practical in large networks due to its worse-case complexity both in terms of messages and computational complexity.

Bonomi et al.~\cite{bonomi2019multi} presented optimizations that improve the performance of Dolev's algorithm on unknown topologies. Indeed, a few modifications allow to save messages, making the algorithm more appealing from a practical point of view even if its worst-case complexity still remains high.
Maurer et al.~\cite{maurer2015communicating} considered the reliable communication problem in settings where the topology can vary with time. In Maurer et al.'s protocol, a process solves the minimum vertex cut problem 
instead of the not equivalent (in dynamic networks) maximum disjoint paths problem. 

\newtext{Several works proposed alternative reliable communication protocols under the local fault model. In particular, Koo presented a broadcast
algorithm under the $t$-locally bounded fault
model~\cite{koo2004broadcast}, which was later coined CPA (Certified
Propagation Algorithm)
by Pelc and Peleg~\cite{pelc2005broadcasting}.}

\newtext{Maurer and Tixeuil have also defined weaker reliable communication primitives for improved scalability~\cite{maurer2014byzantine,maurer2014containing,maurer16scalable}.
CPA was later extended to: \emph{(i)} support different fault thresholds for nodes' neighborhood; \emph{(ii)} consider additional knowledge about the network topology; and \emph{(iii)} consider the general adversary model~\cite{pagourtzis2017reliable}. 
Similarly to aforementioned works~\cite{dolev1981unanimity,bonomi2019multi,maurer2015communicating}, CPA-related solutions also assume a honest dealer.
}

All the aforementioned works~\cite{dolev1981unanimity,bonomi2019multi,maurer2015communicating,litsas2013graph,koo2004broadcast,pelc2005broadcasting} solve a weaker problem than BRB, namely RC. Indeed, they guarantee that all correct processes eventually deliver messages diffused by a correct source but no agreement in case of a faulty one.

More recently, BRB on partially connected networks has been achieved by combining Bracha's and Dolev's algorithms\footnote{One can also combine Bracha and CPA, but the different local Byzantine conditions yields a stronger requirement to be satisfied.}. For example, Wang and Wattenhoffer used this method to design a randomized Byzantine agreement protocol~\cite{wang2020asynchronous}. We show in this work that the two protocols can be combined in a more efficient way. We evaluate the impact of our modifications to the combination of Bracha's algorithm with Bonomi et al.~\cite{bonomi2019multi}'s improved version of Dolev's algorithm.
Our new modifications apply to Bracha's and Dolev's protocols, and the cross-layer combination of the two protocols.

Other approaches have assumed authenticated processes (i.e., processes can use digital signatures) instead of authenticated communication channels~\cite{Castro+Liskov:osdi:pbft:1999}. Relying on cryptography provides integrity and authenticity properties that simplify the algorithms. In particular, weaker connectivity is required to solve the BRB problem. However, cryptography has a computational cost and \newtext{typically} requires a trusted public key infrastructure (PKI). 
From a theoretical point of view, cryptography-based approaches are limited 
to the case of computationally bounded adversaries (e.g., that cannot generate hash collisions). 
Also, having a trusted agent (TA) that is never Byzantine trivializes the BRB problem (the sender can simply send its message to the TA, and every correct node can then reliably collect it from there). By contrast, we aim at solutions that can cope with computationally unbounded adversaries \newtext{and do not make trust asssumptions}.

Recently, Contagion~\cite{guerraoui2019scalable} replaced quorums by smaller stochastic samples, and described an abstraction whose properties can be violated with a probability that depends on the size of the samples. 
Differently, RT-ByzCast~\cite{kozhaya2018rt} and PISTIS~\cite{kozhaya2021pistis} 
aimed at providing real-time guarantees in probabilistically synchronous and reliable networks.
The idea of using pseudo-randomized message dissemination has been used to tolerate malicious or selfish behaviors in several works~\cite{mokhtar2014acting, decouchant2016pag}, which however only provide probabilistic guarantees, while we target exact deterministic guarantees.

\section{System Model and Problem Statement} \label{sec:sysmodel}

\textbf{Processes.} We assume a set $\Pi=\{p_1, p_2, \dots, p_N\}$ of $N$ processes, uniquely identified by an ID. We assume that up to $f < \floor*{N/3}$ of the $N$ processes are Byzantine, i.e., that they can behave arbitrarily or maliciously.  Processes know about the number $N$ of processes, the IDs of the processes in the system, and the fault threshold $f$. 

\begin{figure}[t]
	\centering
	\includegraphics[width=\columnwidth]{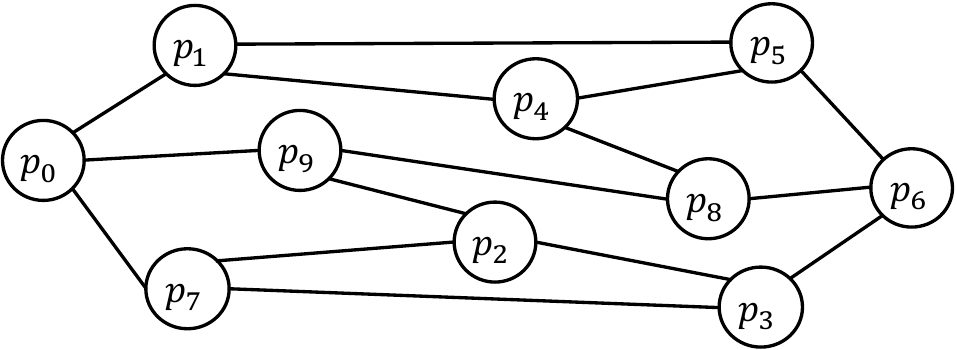}
	\caption{A communication graph with $N{=}10$ nodes and a node connectivity $k{=}3$. \newtext{Any two nodes are either directly connected, or connected through 3 vertex-disjoint paths.}}
	\label{fig:randomgraph}
\end{figure}

\textbf{Communications.} Processes are interconnected by a communication network, which can be represented by an undirected graph $G=(V,E)$ where each node is a process $p_i \in \Pi$ (i.e., $V=\Pi$), and each edge is a communication channel. 
Two processes can directly communicate with each other if and only if they are connected by an edge.
Otherwise, processes have to rely on others to relay their messages and communicate.
Intermediary nodes might be Byzantine, i.e., drop, modify or inject messages. 
We assume that the network topology is unknown to processes. 
Figure~\ref{fig:randomgraph} illustrates a 3-vertex-connected communication network with 10 processes.  

\newtext{
\textbf{Authenticated channels.} 
We assume that communication channels are authenticated, i.e., that messages received by a node $p_i$ on the link interconnecting it with a node $p_j$ always come from $p_j$. Communication channels can be synchronous or asynchronous, but they are reliable, i.e., they do not lose or alter messages. We assume that the communication network is at least $2f{+}1$ vertex-connected so that processes can reliably communicate~\cite{dolev1981unanimity}. 
A node interacts with all its channels through an interface $\mathit{al}$ that exposes two operations, \texttt{Send} and \texttt{Deliver}. A node $p$ can send a message $m$ to a destination $q$ triggering event $\langle\mathit{al},\texttt{Send}\mid{q,m}\rangle$. Similarly, the reception of a message $m$ sent by process $p$ is signaled by event $\langle\mathit{al},\texttt{Deliver}\mid{p,m}\rangle$. 
}

\textbf{Byzantine Reliable Broadcast.}
We consider the Byzantine Reliable Broadcast (BRB) abstraction, which guarantees the following properties:

\noindent \textbf{[BRB-Validity]} If a correct process $p$ broadcasts $m$,
then every correct process eventually delivers $m$.

\noindent \textbf{[BRB-No duplication]} No correct process delivers
message $m$ more than once.

\noindent \textbf{[BRB-Integrity]} If a correct process delivers a message
$m$ from sender $p_i$, then $m$ was previously broadcast by
$p_i$.

\noindent \textbf{[BRB-Agreement]} If some correct process
  delivers $m$, then every correct process eventually
  delivers~$m$. 
  
We assume that processes broadcast payload data, which can be of variable size, and
that they might have to broadcast the same payload data at different times during the system's life. To do so, processes use a header that uniquely identifies the payload data and contains control information. 

\newtext{
A node initiates the broadcast of a message $m$ triggering event
 $\langle Bracha, \texttt{Broadcast}\ |\ m \rangle$, while a $\langle Bracha, \texttt{Deliver}\ |\ s, m \rangle$ event indicates
that a message $m$ broadcast by node $s$ can be delivered.}

\textbf{Reliable Communication.}
In addition to the BRB abstraction, we also consider the weaker
Reliable Communication (RC) abstraction, which has the same interface
as BRB, and guarantees the same properties, except BRB-Agreement,
which is only guaranteed when the sender is correct.
\newtext{
The interface of RC is defined as follows: nodes initiate the broadcast of a message $m$, which might include the actual payload data and some additional information,  by triggering event
 $\langle Dolev, \texttt{Broadcast}\ |\ m \rangle$, while a $\langle Dolev, \texttt{Deliver}\ |\ m \rangle$ event indicates
that a message $m$ can be delivered. Note that the $\texttt{Deliver}$ event does not indicate the source of a message, while the original RC protocol presented by Dolev does. This comes from the fact that we base our RC layer on Bonomi et al.'s protocol~\cite{bonomi2019multi}, which improves the performance of Dolev's protocol but sometimes prevents the identification of the source of a message based solely on paths and therefore requires the source of a broadcast message to be included in the payload data.}

\section{Background: BRB on asynchronous and Partially Connected Networks}
\label{sec:background}

In this section we first recall Bracha's and Dolev's algorithms and we then present how they can be used to provide BRB in asynchronous partially connected networks. 

\subsection{BRB in Asynchronous and Fully Connected Networks}

Bracha's algorithm (sometimes called \emph{authenticated double-echo broadcast}) describes the first BRB protocol for asynchronous and authenticated communication networks~\cite{Bracha:iandc:1987}. 
Let us recall that this protocol assumes a fully connected network made of reliable and authenticated point-to-point links and it tolerates up to $f$ Byzantine processes (where $f < N/3$ and $N$ is the number of processes in the network).  

The protocol works in three phases. When a process $p_i$ wants to BRB-broadcast some payload data, it sends a \texttt{SEND}
message along with the payload data to all nodes in the system (phase 1).
Upon receiving a \texttt{SEND} message, a process sends an \texttt{ECHO}
message to all the nodes in the system (along with the payload data) and moves to phase 2 where it remains waiting for a quorum of \texttt{ECHO} messages.
Upon receiving $\ceil*{\frac{N+f{+}1}{2}}$ \texttt{ECHO} messages for a given payload, a process sends a \texttt{READY} message to other
nodes and moves to phase 3.
Let us note that a \texttt{READY} message can also be sent upon receiving $f{+}1$ \texttt{READY}
messages as it will ensure that al least one correct process is moving to phase 3 and thus it is safe to broadcast a \texttt{READY} message and to move to phase 3 as well. 
Finally, upon receiving $2f{+}1$ \texttt{READY} messages,
a process can BRB-deliver the payload data by concluding the procedure.

For the sake of completeness, we report in Algorithm~\ref{alg:bracha} the pseudo-code of the algorithm, which we extract from~\cite{Cachin+Guerraoui+Rodrigues:2011, raynal2018fault}.
\newtext{This protocol has been proven to satisfy the
  \textit{BRB-validity}, \textit{BRB-no duplication},
  \textit{BRB-integrity} and \textit{BRB-agreement} properties.}
  
\begin{algorithm}[t]
\caption{ \textit{BRB in asynchronous and fully connected networks (Bracha's protocol) at process $p_i$}} \label{alg:bracha}
\begin{algorithmic}[1]
\footnotesize 	
\State \textbf{Parameters:}
\State \hspace*{2em} $\Pi$: the set of all processes.
\State \hspace*{2em} $N$: total number of processes. 
\State \hspace*{2em} $f < N/3$: maximum number of Byzantine processes.
\footnotesize 	\State \textbf{Uses:}  Auth. async. perfect point-to-point links, instance \textit{al}. \newtext{al is a channel or a set of channels?}
\State

\State \textbf{upon event} $\langle Bracha, \texttt{Init} \rangle$ \textbf{do}
    \State \hspace*{2em} $sentEcho = sentReady = delivered = \textbf{False}$
    \State \hspace*{2em} $echos = readys = \emptyset$ 
\State

\State \textbf{upon event} $\langle Bracha, \texttt{Broadcast}\ |\ m \rangle$ \textbf{do}
    \State \hspace*{2em} \textbf{forall} $q \in \Pi$ \textbf{do} $\{$ \textbf{trigger} $\langle al, \texttt{Send}\ |\ q, [\texttt{SEND}, m] \rangle$ $\}$\label{alg:bracha-send-send}
\State

\State \textbf{upon event} $\langle al, \texttt{Deliver}\ |\ p, [\texttt{SEND}, m] \rangle$ \textbf{and not} $sentEcho$ \textbf{do}\label{alg:bracha-deliver-send} 
    \State \hspace*{2em} $sentEcho = \textbf{True}$
    \State \hspace*{2em} \textbf{forall} $q \in \Pi$ \textbf{do} $\{$ \textbf{trigger} $\langle al, \texttt{Send}\ |\ q, [\texttt{ECHO}, m] \rangle$ $\}$\label{alg:bracha-send-echo}
\State

\State \textbf{upon event} $\langle al, \texttt{Deliver}\ |\ p, [\texttt{ECHO}, m] \rangle$ \textbf{do}\label{alg:bracha-deliver-echo}
    \State \hspace*{2em}
 $echos.\texttt{insert}(p)$
\State

\State \textbf{upon event} $echos.\texttt{size}() \ge \ceil{\frac{N+f{+}1}{2}}$ \textbf{and not} $sentReady$ \textbf{do}
    \State \hspace*{2em} $sentReady = \textbf{True}$
    \State \hspace*{2em} \textbf{forall} $q \in \Pi$ \textbf{do} $\{$ \textbf{trigger} $\langle al, \texttt{Send}\ |\ q, [\texttt{READY}, m] \rangle$ $\}$\label{alg:bracha-send-ready}
\State

\State \textbf{upon event} $\langle al, \texttt{Deliver}\ |\ p, [\texttt{READY}, m] \rangle$ \textbf{do}\label{alg:bracha-deliver-ready}
    \State \hspace*{2em}
 $readys.\texttt{insert}(p)$
\State

\State \textbf{upon event} $readys.\texttt{size}() \ge f{+}1$ \textbf{and not} $sentReady$ \textbf{do} 
    \State \hspace*{2em} $sentReady = \textbf{True}$
    \State \hspace*{2em} \textbf{forall} $q \in \Pi$ \textbf{do} $\{$ \textbf{trigger} $\langle al, \texttt{Send}\ |\ q, [\texttt{READY}, m] \rangle$ $\}$\label{alg:bracha-send-ready2}
\State

\State \textbf{upon event} $readys.\texttt{size}() \ge 2f{+}1$ \textbf{and not} $delivered$ \textbf{do} 
    \State \hspace*{2em} $delivered = \textbf{True}$
    \State \hspace*{2em} \textbf{trigger} $\langle Bracha, \texttt{Deliver}\ |\ s, m \rangle$ 
\end{algorithmic}
\end{algorithm}

\subsection{Reliable Communication in ($2f\!+\!1$)-Connected Networks in the Global Fault Model}

Dolev's protocol provides reliable communication if processes are interconnected by reliable and authenticated communication channels,
and if the communication network is at least $(2f{+}1)$-connected~\cite{dolev1981unanimity}.

Dolev's protocol leverages the authenticated channels to collect the
label of processes traversed by a content.
Processes use those labels to compute the maximum number of node-disjoint
paths among all the paths traversed for a particular content.
A process delivers a content as soon as it has verified its
authenticity, i.e., as soon as it has received the content
through at least $f{+}1$ node-disjoint paths.
The keystone of Dolev's proof is that by Menger's theorem,
if the communication network is $(2f{+}1)$-connected then there are at
least $2f{+}1$ node-disjoint paths between any two processes in the
network~\cite{menger1927allgemeinen}, and since at most $f$
processes are faulty, according to the pigeonhole principle, a process
will receive a message sent by another process through at least $f{+}1$
disjoint paths.

Depending on whether the processes know the network topology or not, Dolev presented two variants of its protocol, which respectively use predefined routes between processes, or flooding. We focus on the unknown topology version of Dolev's algorithm, whose pseudo-code is presented in Algorithm~\ref{alg:dolev}. 

\textbf{Dolev's algorithm made practical.}
\label{sec:bonomi_optims}
Let us note that the worst case complexity of Dolev's algorithm is high (both in terms of number of messages and complexity to verify if a specific content can be safely delivered).
Bonomi et al. presented several modifications that reduce the number of messages transmitted along with their size in practical executions~\cite{bonomi2019multi}.
We recall these modifications in the following.

\noindent \textbf{[MD.1]} If a process $p$ receives a content directly from the source $s$, then $p$ directly delivers it.

\noindent \textbf{[MD.2]} If a process $p$ has delivered a content, then it can discard all the related paths and relay the content only with an empty path to all of its neighbors.

\noindent \textbf{[MD.3]} A process $p$ relays path related to a content only to the neighbors that have not delivered it.

\noindent \textbf{[MD.4]} If a process $p$ receives a content with an empty path from a neighbor $q$, then $p$ can abstain from relaying and analyzing any further path related to the content that contains the label of $q$.

\noindent \textbf{[MD.5]} A process $p$ stops relaying further paths related to a content after it has been delivered and the empty path has been forwarded. 

\newtext{Let us stress that modifications MD.1--5 have not been proven to improve the worst case theoretical complexity of the protocol, but the authors showed, through experimental evaluations that their performance gain is significant in several practical settings.} 

\begin{algorithm}[t]
\caption{ \textit{Reliable communication in $(2f{+}1)$-connected networks (Dolev's protocol) at process $p_i$}} \label{alg:dolev}
\begin{algorithmic}[1]
\footnotesize 	\State \textbf{Parameters:}
\State \hspace*{2em} $f$ : max. number of Byzantine processes in the system.
\footnotesize 	\State \textbf{Uses:} Auth. async. perfect point-to-point links, instance \textit{al}. 
\State

\Event{$\langle Dolev, \texttt{Init} \rangle$}
\State $delivered = \textbf{False}$
\State $paths = \emptyset$
\EndEvent

\State \textbf{upon event} $\langle Dolev, \texttt{Broadcast}\ |\ m \rangle$ \textbf{do}
    \State \hspace*{2em} \textbf{forall} $p_j \in \texttt{neighbors}(p_i)$ \textbf{do}
        \State \hspace*{4em} \textbf{trigger} $\langle al, \texttt{Send}\ |\ p_j, [m, [\ ]] \rangle$
    \State \hspace*{2em} $delivered = \textbf{True}$
    \State \hspace*{2em} \textbf{trigger} $\langle Dolev, \texttt{Deliver}\ |\ m \rangle$
\State

\State \textbf{upon event} $\langle al, \texttt{Deliver}\ |\ p_j, [m, path] \rangle$ \textbf{do}
    \State \hspace*{2em}  $paths.\texttt{insert}(path + [p_j])$
    \State \hspace*{2em} \textbf{forall} $p_k \in \texttt{neighbors}(p_i) \setminus (path \cup \{p_j\})$ \textbf{do}
    \State \hspace*{4em} \textbf{trigger} $\langle al, \texttt{Send}\ |\ p_k, [m, path + [p_j]] \rangle$
\State

\State \textbf{upon event} ($p_i$ is connected to the source through $f{+}1$ node-disjoint paths contained in $paths$) \textbf{and} $delivered = \texttt{False}$ \textbf{do}\label{alg:dolev-disj-paths}
        \State \hspace*{4em} \textbf{trigger} $\langle Dolev, \texttt{Deliver}\ |\ m \rangle$ 
        \State \hspace*{4em} $delivered = \textbf{True}$
\end{algorithmic}
\end{algorithm}

\subsection{BRB in partially connected networks}

The state-of-the-art method to implement the BRB abstraction in a partially connected network consists in combining Bracha's algorithm with a second algorithm that is in charge of providing the abstraction of a reliable point-to-point link (i.e., with a protocol ensuring Reliable Communications on generic networks). 
For that purpose, one can either use Dolev's algorithm, CPA, or even topology specific protocols~\cite{behrens2018rdmc}, depending on the assumptions one is willing to make regarding the communication network. 

In the global fault model and without additional network assumptions, one can rely on Dolev's reliable communication protocol. 
The combination of Bracha's and Dolev's algorithms has recently been used to design a randomized Byzantine agreement protocol on partially connected networks~\cite{wang2020asynchronous}. 

In practice, to combine these protocols one can replace
each send-to-all operation at process $p_i$ (i.e, \textbf{forall} $q\!\in\!\Pi$ \textbf{do} $\{ \textbf{trigger} \langle al, \texttt{Send}\, |\, q, [mType, m] \rangle \}$) in Algorithm~\ref{alg:bracha}
(lns.~\ref{alg:bracha-send-send},~\ref{alg:bracha-send-echo},~\ref{alg:bracha-send-ready},~\ref{alg:bracha-send-ready2}) by a $\langle Dolev, \texttt{Broadcast}\, |\, [mType, m]  \rangle$
operation, where $mType\! \in\! \{\texttt{SEND}, \texttt{ECHO}, \texttt{READY}\}$.

In addition, one has to report the ID of the source of a delivered message in Dolev, and trigger the code triggered by $\langle\mathit{al},\texttt{Deliver}\mid{q},[msgType, m]\rangle$ in Algorithm~\ref{alg:bracha} (lns.~\ref{alg:bracha-deliver-send},~\ref{alg:bracha-deliver-echo},~\ref{alg:bracha-deliver-ready}) whenever $\langle\mathit{Dolev},\texttt{Deliver}\mid[source, msgType, m] \rangle$ is executed following the identification of $f{+}1$ disjoint paths to $source$ have been identified in Algorithm~\ref{alg:dolev}, ln.~\ref{alg:dolev-disj-paths}.

\begin{figure}[t]
	\centering
	\includegraphics[width=.9\columnwidth]{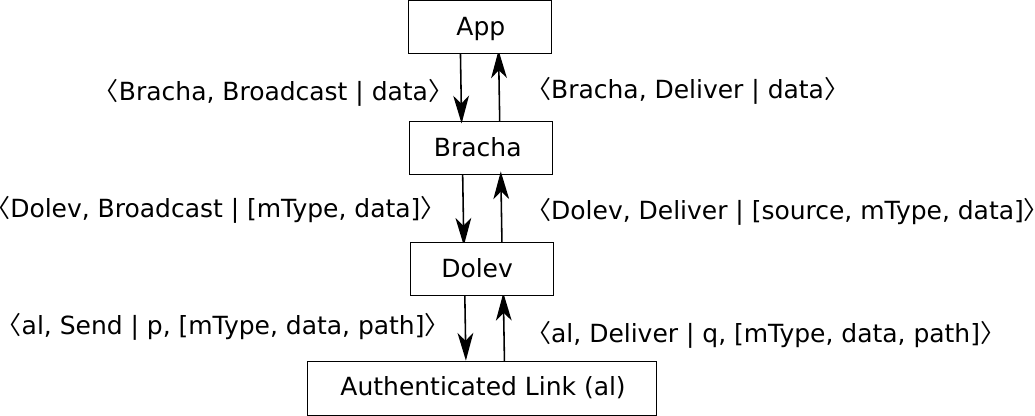}
	\caption{Composition of Bracha's and Dolev's algorithms to implement BRB on partially connected networks.}
	\label{fig:interfacesBrachaDolev}
\end{figure}

The resulting protocol stack is illustrated in Figure~\ref{fig:interfacesBrachaDolev}, which also shows the interface of each module. 
This combination can be made more practical by applying modifications MD.1--5 \newtext{to the Dolev RC layer.} 
However, given that the number of messages generated by Bracha's and Dolev's protocols are respectively $\mathcal{O}(N^2)$ and $\mathcal{O}(N!)$, and that Dolev's algorithm requires solving a problem with exponential complexity (i.e., finding disjoint paths in a unknown network),
the combination of  these two protocols does not scale well with the number of processes.

\section{Cross-layer Bracha-Dolev Implementation and Functional Modifications} \label{sec:brachadolev}

\begin{figure}[t]
	\centering
	\includegraphics[width=.8\columnwidth]{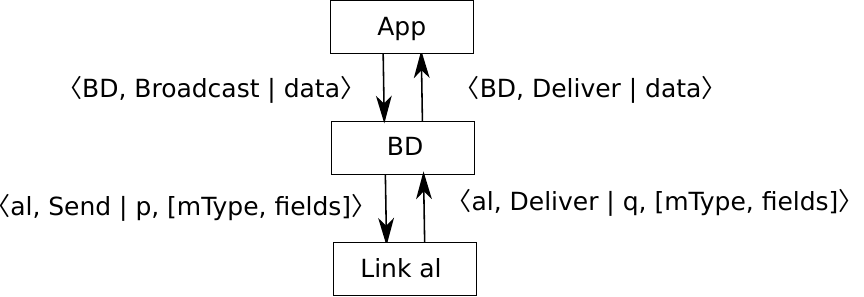}
	\caption{Interfaces in a cross-layer combination of Bracha's and Dolev's protocols to implement BRB on partially connected networks.}
	\label{fig:interfacesBrachaDolev_modified}
\end{figure}

\textbf{Protocol stack and interfaces.}
Figure~\ref{fig:interfacesBrachaDolev_modified} illustrates the
interface of our combination of Bracha's and Dolev's protocols,
modified to support modifications MD.1--5
and repeatable broadcasts. We detail below the fields used in this
interface, and Sec.~\ref{sec:fields} shows how some of those fields can be made optional. 

\textbf{Repeatable broadcast.}
Across the system's life, we assume that a process might broadcast
several times the same payload data. For example, this 
would be required for sensing applications (e.g., temperature
monitoring). 
We then assume that a
broadcast message contains the payload data $m$, the ID of its source
process $s$, and a broadcast ID $bid$ (a sequence number) that the source process
monotonically increases after each broadcast. If the source is
correct, the source and broadcast IDs of a broadcast message
uniquely identify a payload data. A Byzantine process might reuse a
broadcast ID for several payload data, in which case all processes will
agree on delivering at most one payload and ignore the others.

We therefore modify the interface of the BRB protocol so that Broadcast and Deliver operations include the $s$ and $bid$ fields: $\langle \mathit{BD}, \texttt{Broadcast}\, |\, [(s, bid), m] \rangle$ and $\langle \mathit{BD}, \texttt{Deliver}\, |\, [(s, bid), m] \rangle$. This modification limits the Byzantine processes' ability to replay messages. 

\textbf{Applying modifications MD.1--5.} For MD.1--5 to be used in the combination of Bracha's and Dolev's algorithms, the format of the ECHO and READY messages has to be slightly modified to contain the ID of their original sender, so that ECHO or READY messages that are broadcast by different processes can be distinguished. Using the original specification of Dolev's protocol, one could rely on the full paths received along with messages to identify the sender of an ECHO or READY message. However, optimization MD.2 of Dolev's protocol replaces a path by an empty path upon delivery of a message, which then prevents the identification of the original sender of the message from its path. For this reason, ECHO and READY messages must contain an additional field that identifies their sender. For example, an ECHO message generated by process~$p_i$ would then have the format $[\texttt{ECHO}, p_i, (s,bid), m, path]$. 

\newtext{
Table~\ref{tab:summary_modifications} in Appx.~\ref{sec:implemDetails} recalls MD.1--5, the state-of-the-art modifications of Dolev's protocol~\cite{bonomi2019multi}, and the novel modifications MBD.1-12 that we detail in the next sections.}

\section{Optimizing for Latency and Throughput}
\label{sec:latency}

\subsection{Limiting payload transmissions} 

Bracha's and Dolev's protocols make processes include the payload data (i.e., the content they wish to broadcast) in each message sent to their neighbors. However, when the payload data is large, and because of the large number of messages generated, it is worthwhile to reduce the number of times the payload data is exchanged. 
Some Byzantine-resilient protocols replace the payload by hashes or signatures for this purpose. These methods could be used in the Bracha-Dolev combination too. However, we aim at avoiding the use of cryptographic methods to contain the computational overhead of the protocol but also to tolerate computationally unbounded adversaries. We modify the protocol as follows. 

\textbf{Payload to local ID association [MBD.1].} 
Upon receiving a message that contains a previously unknown payload data \newtext{from a given source and with a given broadcast ID}, a process~$p_i$ associates it to a novel locally-generated and unique ID that will only be used with its direct neighbors.
When sending a message to its neighbors that is related to this payload data for the first time, the process then includes the local ID it chooses. In later messages, $p_i$ only sends the local ID of that payload data.
A neighbor of $p_i$, say $p_j$, might generate its own local ID for a given payload before receiving $p_i$'s local ID. \newtext{A process should however always use its own local ID to refer to a given payload data to avoid collisions with the IDs chosen by its neighbors.}  

Because of the asynchrony of communications, a process might actually receive the message that contains the payload data after subsequent messages that only use the local ID. In that case, we modified the protocol so that a process stores the messages that mention an unknown payload in a queue, and processes them all when the payload data is finally received. 
In the worst case, our method requires a process to receive a given payload data once from each of its neighbors. In practice, the size of local IDs should be decided in an ad-hoc manner so that a process does not run out of available IDs.   

\subsection{Bracha phase transitions}

Processes receive and forward Send, Echo and Ready messages that have been created according to Bracha's protocol, following a dissemination that Dolev's protocol determines.
In the following, we describe several modifications that processes can implement when they create messages, following the reception of a message that has been delivered using Dolev's algorithm.
These modifications reduce the total number of messages exchanged, and the overall bandwidth consumption. 

\textbf{Echo and Ready amplifications.}
In Bracha's protocol, it is known (and necessary) that the reception of $f{+}1$ Readys allows a process to generate its own Ready message, if it has not done so already~\cite{Cachin+Guerraoui+Rodrigues:2011, raynal2018fault}. This situation can happen in Bracha's protocol because the links are asynchronous. However, we also observe that the delivery of $f{+}1$ Echos allows a process to generate its own Echo message, and send it if it was not sent before. In addition, our modification MBD.2 of Send messages into single-hop messages makes this amplifications of Echos necessary. 
\newtext{To accelerate the creation of an Echo message, one can also observe that delivering a Ready message allows the immediate creation and delivery of the associated Echo message.}
\newtext{It might also happen that the Dolev-delivery of an Echo message allows a process to generate both its own Echo and its own Ready message. In such a case, the process only sends a Ready message.}

\textbf{Single-hop Send messages [MBD.2].}
In the default combination of Bracha's and Dolev's protocols,
the Send messages that a correct process creates reach all the processes. We modify the protocol so that upon receiving a Send message, a neighbor of the source stops disseminating it, and creates instead an Echo message that it relays to all its neighbors.
When a process $p_i$ Dolev-delivers a $[\texttt{SEND}, (s, bid), m,
  path]$ message that it received from a neighbor $p_j$, the
combination of Bracha's and Dolev's algorithms makes it forward a
modified Send message to its neighbors not included in $path$, i.e.,
$[\texttt{SEND}, (s, bid), m, path{+}[p_j]]$.
A process can avoid generating and forwarding the modified Send
message without loss of information, \newtext{as it can be extracted from Echo messages}. A Send message is therefore a single-hop message and does not need to carry a $path$. \newtext{More generally, note that an Echo message from the source can be extracted from any message that a process receives, which might also enable a node to send its own Echo or Ready faster.}
\begin{proof}
Bonomi et al. proved that upon Dolev-delivery of a message $[\texttt{SEND}, (s, bid), m, path]$ a process $p_i$ can send a $[\texttt{SEND}, (s, bid), m, [\,]\,]$ message instead of the $[\texttt{SEND}, (s, bid), m, path\! +\! [p_j]\,]$ message to its neighbors not included in $path$ (MD.2 in Sec.~\ref{sec:bonomi_optims}). This modification decreases the size of the forwarded Send message. 
Following Bracha's algorithm, after having validated a Send message, process $p_i$ should also send an $[\texttt{ECHO}, p_i, (s, bid), m, [\,]\,]$ message to all its neighbors. 
However, upon receiving an $[\texttt{ECHO}, p_i, (s, bid), m, path]$ message, processes can extract the $[\texttt{SEND}, (s, bid), m, path]$ message they should have received in the unmodified protocol and process it.\\
\end{proof} 

\textbf{Echo to Echo transitions [MBD.3].} When a process $p_i$ receives an $[\texttt{ECHO},p_k,(s,bid),m,path]$ from a neighbor $p_j$ that makes $p_i$ Dolev-deliver the message, according to Dolev's protocol and MD.2, $p_i$ forwards $[\texttt{ECHO},p_k,(s,bid),m,[\,]\,]$ to its neighbors not included in $\mathit{path}$. As a result of delivering the Echo message, process $p_i$ might also send an $[\texttt{ECHO}, p_i, (s, bid), m, [\,]\,]$ message to all its neighbors (after having delivered the Send message of the source, or received $f{+}1$ Echos - using the \textit{Echo amplification}). 
These two messages can be merged into a single one, in particular because they both have to be transmitted using empty paths (the first one because it was delivered, due to MD.2, and the second one because it has just been created). For that purpose, we introduce a new message type Echo\_Echo formatted as $[\texttt{ECHO\_ECHO}, p_i, p_k, (s, bid), m, path]$, where $p_k$ is the ID of the process whose Echo message was received, while $p_i$ is the ID of the process who Dolev-delivered the Echo message and is sending its own Echo.
We explain below how $\texttt{ECHO\_ECHO}$ messages are handled upon reception.

Finally, a process $p_i$ can decide to which neighbors it should send
the Echo\_Echo message, and to which it should only
send its own Echo message (cf. MD.3,~\ref{sec:bonomi_optims}), \newtext{or no messages at all (cf. MBD.8)}: (i)
$[\texttt{ECHO\_ECHO}, p_i, p_k, (s, bid), m, [\,]\,]$ is sent
to the neighbors that have not yet Dolev-delivered
$[\texttt{ECHO}, p_k, (s, bid), m]$ and are not included in
the received path; \newtext{(ii) $[\texttt{ECHO}, p_i, (s, bid), m, [\,]\,]$ is sent to the remaining neighbors that did not BD-deliver.}

\begin{proof}
Upon receiving an $[\texttt{ECHO\_ECHO}, p_i, p_k,$ $(s, bid), m, [\,]\,]$ message, a process can extract the $[\texttt{ECHO}, p_i, (s, bid), m, [\,]\,]$ and $[\texttt{ECHO}, p_k, (s, bid), m, [\,]\,]$ messages. It is correct for~$p_i$ to forward the Echo message with an empty path (MD.2,~\ref{sec:bonomi_optims}). In addition, $p_i$ does not have to forward the $[\texttt{ECHO}, p_k, (s, bid), m, [\,]\,]$ message to its neighbors that have delivered the $[\texttt{ECHO}, p_k, (s, bid), m]$ message (MD.3,~\ref{sec:bonomi_optims}), which justifies the use of an Echo message instead of an Echo\_Echo message. \newtext{Finally, $p_i$ does not have to send any message to a neighbor that delivered the content, i.e., relayed $2f{+}1$ Readys with empty paths, as will be shown in MBD.8.} \\
\end{proof}

\textbf{Echo to Ready transitions [MBD.4].} When a process~$p_i$ Dolev-delivers an $[\texttt{ECHO}, p_j, (s, bid), m, path]$ message that it received from a neighbor $p_k$, the standard Bracha-Dolev protocol combination makes $p_i$ forward a modified Echo message to its neighbors not included in $path$, i.e., $[\texttt{ECHO}, p_j, (s, bid), m, path\! +\! [p_k]\,]$, while MD.2 makes $p_i$ forward an empty path to those neighbors.
As a result of delivering the Echo message, process $p_i$ might also send a $[\texttt{READY}, p_i, (s, bid), m, [\,]\,]$ message to all its neighbors (after having Dolev-delivered $2f{+}1$ Echo messages).  

We introduce a second novel message type, Ready\_Echo, associated to messages of the format $[\texttt{READY\_ECHO}, p_i, p_j, (s, bid), m, path]$,
where $p_i$ is the ID of the process whose Ready is received, while $p_j$ is the ID of the process whose Echo was Dolev-delivered by $p_i$ and triggered the emission of $p_i$'s Ready message. The $path$ field of a Ready\_Echo message describes 
an empty path upon creation, but this is not the case if the message is relayed.  
As for Echo\_Echo messages, process $p_i$ decides whether to send a Ready\_Echo, Ready, Echo, or no message at all for each of its neighbors depending on whether or not it has previously transmitted an empty path for the Echo or the Ready.

\begin{proof} 
Upon receiving a $[\texttt{READY\_ECHO}, p_i, p_j,$ $(s, bid), m, [\,]\,]$ message, a process can extract $[\texttt{ECHO}, p_j, (s, bid), m, [\,]\,]$ and $[\texttt{READY}, p_i, (s, bid), m, [\,]\,]$ messages. It is correct for $p_i$ to forward the Echo message with an empty path (MD.2,~\ref{sec:bonomi_optims}). In addition, $p_i$ does not have to forward the $[\texttt{ECHO}, p_j, (s, bid), m, [\,]\,]$ message to its neighbors that have delivered the $[\texttt{ECHO}, p_j, (s, bid), m]$ message (MD.3,~\ref{sec:bonomi_optims}), which justifies the use of a Ready message instead of a Ready\_Echo message for those neighbors. \\
\end{proof}

\begin{algorithm}[t]
\caption{ \textit{$\texttt{READY\_ECHO}$ message reception at process $p_i$}} \label{alg:ready2}
\begin{algorithmic}[1]
\footnotesize 
\Event{$\langle$ $al$, $\texttt{Deliver}$ | $p_l$, [$\texttt{READY\_ECHO}$, $p_r$, $p_e$, (s, bid), m, path] $\rangle$ }
    \State $\texttt{insert}$ $path\! +\! [p_l]$ in $[\texttt{ECHO}, p_e, (s, bid)]$'s graph
    \State $\texttt{insert}$ $path\! +\! [p_l,p_e]$ in $[\texttt{READY}, p_r, (s, bid)]$'s graph \State

    \State \comalgo{// $\texttt{Dolev-delivered}(\mathit{msg})$ checks whether $\mathit{msg}$ has already been}
    \State \comalgo{// Dolev-delivered prior to this event (MD.5)}
    \State $sendEcho\!=\, !\texttt{Dolev-delivered}([\texttt{ECHO}, p_e, (s, bid)])$
    \State $sendReady\!=\, !\texttt{Dolev-delivered}([\texttt{READY}, p_r, (s, bid)])$ 
    \State
    \State \comalgo{// $\texttt{actOnEchos}$ and $\texttt{actOnReadys}$ create and send Echo/Ready}
    \State \comalgo{// messages if necessary, according to Bracha's algorithm}
    \State \comalgo{// $\texttt{Dolev-delivering}(\mathit{msg})$ checks whether $\mathit{msg}$ is being}
    \State \comalgo{// Dolev-delivered at this event (MD.2)}
    \State ($\texttt{Dolev-delivering}([\texttt{ECHO}, p_e, (s, bid)]))?$ $\{\texttt{actOnEchos}(s, bid, p_e);$ $epath\! =\! [\,]\}:$ $epath\! =\! path\! +\! [p_l]$ 
    \State ($\texttt{Dolev-delivering}([\texttt{READY}, p_r, (s, bid)]))?$ $\{\texttt{actOnReadys}(s, bid, p_r);$ $rpath\!=\![\,]\}:$ $rpath\! =\! path\! +\! [p_l]$ 
    \State
    
    \State \comalgo{// Avoiding the neighbors that sent empty paths (MD.3)}
    \State $nde = \texttt{neighborsThatDelivered}([\texttt{ECHO}, p_e, (s, bid)])$
    \State $ndr = \texttt{neighborsThatDelivered}([\texttt{READY}, p_r, (s, bid)])$ 
    \State
    
    \State \comalgo{// Avoiding the neighbors that Bracha-delivered (MBD.8)}
    \State $nd = \texttt{neighborsThatDelivered}((s, bid))$ 
    \State
            
    \If{$sendEcho$ \textbf{and} $!sendReady$}
        \State \textbf{forall} $x \in \texttt{neighbors}(p_i) \setminus (nd\! \cup\!nde\! \cup\! path\! \cup\! \{p_l\})$
            \State \hspace*{5mm} \textbf{trigger} $\langle{al},\texttt{Send}\mid{x},[\texttt{ECHO},p_e,(s, bid),m,epath])\rangle$
            
    \ElsIf{$!sendEcho$ \textbf{and} $sendReady$}

        \State \textbf{forall} $x \in \texttt{neighbors}(p_i) \setminus (nd\! \cup\!ndr\! \cup\! path\! \cup\! \{p_l\})$
            \State \hspace*{5mm} \textbf{trigger} $\langle{al},\texttt{Send}\mid{x},[\texttt{READY},p_r,(s, bid),m,rpath])\rangle$
    \Else
        \State \textbf{forall} $x \in ndr \setminus (nd\! \cup\! nde\! \cup\! path\! \cup\! \{p_l\})$
            \State \hspace*{5mm} \textbf{trigger} $\langle{al},\texttt{Send}\mid{x},[\texttt{ECHO},p_r,(s, bid),m,rpath])\rangle$
        \State \textbf{forall} $x \in nde \setminus (nd\! \cup\! ndr\! \cup\! path\! \cup\! \{p_l\})$
            \State \hspace*{5mm} \textbf{trigger} $\langle{al},\texttt{Send}\mid{x},[\texttt{READY},p_r,(s, bid),m,rpath])\rangle$
        \State \textbf{forall} $x \in neighbors(p_i) \setminus (nd\! \cup\! nde\! \cup\! ndr\! \cup\! path\! \cup\! \{p_l\})$
            \State \hspace*{5mm} {\textbf{if} $epath == rpath$ \textbf{then}}
                \State \hspace*{5mm}\hspace*{5mm} $\textbf{trigger}\, \langle{al},\, \texttt{Send}\, |\, x,\! [\texttt{READY\_ECHO},\!$ $p_e,\!$ $p_r,\!$ $(s, bid),\! m,\! epath]\rangle$
            \State \hspace*{5mm} {\textbf{else}}
                \State \hspace*{5mm}\hspace*{5mm} $\textbf{trigger}\, \langle{al},\texttt{Send}\mid{x},[\texttt{ECHO},p_e,(s,bid),m,epath]\rangle$
                \State \hspace*{5mm}\hspace*{5mm} $\textbf{trigger}\, \langle{al},\!\texttt{Send}\mid{x},[\texttt{READY},p_r,(s,\!bid),m,rpath]\rangle$
            \State \hspace*{5mm} \textbf{end if}
    \EndIf
\EndEvent
\end{algorithmic}
\end{algorithm}

\textbf{Reception of Echo\_Echo and Ready\_Echo messages.} We detail
here how processes handle Ready\_Echo messages (Echo\_Echos are handled
similarly). Upon receiving a $[\texttt{READY\_ECHO}, p_i, p_j, (s,
  bid), m, path]$ message, we have seen that processes can extract
$[\texttt{ECHO}, p_j, (s, bid), m, path]$ and $[\texttt{READY}, p_i,
  (s, bid), m, path]$ messages. However, it might not always be
necessary for processes to forward the Ready\_Echo message to their
neighbors, as a Dolev dissemination would dictate. Indeed, some of
these neighbors might have already delivered the Echo or the Ready
message (MD.3,~\ref{sec:bonomi_optims}), or neighbors that have
delivered can be included in the received path
(MD.4,~\ref{sec:bonomi_optims}). Algorithm~\ref{alg:ready2} provides
the pseudocode that processes use to handle the Ready\_Echo messages they receive. 

\textbf{Empirical lessons.} We observed that the system's latency is lower when processes first forward a message they have received before sending the message they might have created. 
In addition, one might also remark that we have not mentioned the possibility for a received Ready message to trigger the emission of a Ready message, which might be exploited to create a Ready\_Ready message. Similarly, one could imagine that receiving an Echo\_Echo message might trigger the emission of an Echo message, which would justify the need to create a message type that would carry three Echos. In practice, we have not observed that these phenomena are rare enough to justify not handling them with specific code.

\subsection{Additional message types for optional message fields [MBD.5]}
\label{sec:fields}

So far, we have assumed that messages in the Bracha-Dolev protocol combination contain, besides the payload data, a payload ID $(s, bid)$, a message type $mType \in \{\texttt{SEND}, \texttt{ECHO}, \texttt{READY}\}$, the ID of the process that generated the message, and a path that indicates the list of process IDs the message went through in the topology. 
We now detail several observations that allow us to distinguish multiple particular cases using different message types, and transmit equivalent information more concisely. 

\textbf{Send messages.} As we have seen in the previous section, Send messages are single-hop messages. 
Therefore, they do not have to carry the source ID in the payload's ID $(s, bid)$, because communication links are authenticated.
Send messages also always carry the payload data along with the local ID that was chosen by the source. \newtext{With MBD.1, Send messages are formatted as $[\texttt{SEND}, bid, \mathit{localPayloadID}, \mathit{payloadSize}, \mathit{payload}]$,  while without MBD.1 they are formatted as $[\texttt{SEND}, bid, \mathit{payloadSize}, \mathit{payload}]$.
}

\newtext{\textbf{Removal of the sender field.} Newly generated Echo or Ready messages are transmitted to neighboring processes without a sender field, since this information is also indicated to the receiving process by the authenticated channel a message is received with. Upon reception, these messages are disseminated with a sender field, but the sender is not added to the dissemination path as it would be redundant.}

\newtext{\textbf{Messages with a local ID and no payload data.} As a consequence to MBD.1, some messages are transmitted without the $payloadSize$, $payload$, $s$ and $bId$ fields, because they can be inferred by the receiving nodes using the $localId$ that is transmitted instead.}

\subsection{Handling asynchrony}

We present several optimizations that can be implemented when the Bracha part of the protocol generates novel messages while the Dolev propagation of a received message has to continue. These optimizations reduce the overall amount of information exchanged over the network. 

\textbf{Ignore Echos received after Dolev-delivering the corresponding Ready [MBD.6].}
If a process $p_i$ Dolev-delivers a Ready message that was sent by a process $p_j$ then $p_i$ can stop disseminating and discard any Echo message emitted by $p_j$.

\begin{proof}
If $p_i$ Dolev-delivered the Ready message of a process~$p_j$, it means that $p_i$ verified that $p_j$ did send the Ready message.
The Echo message that $p_j$ sent contains less information and reflects an old state of $p_j$.  
\end{proof}

\textbf{Ignore Echos received after delivering the content [MBD.7].}
If a process $p_i$ Bracha-delivers a content (because it has Dolev-delivered $2f{+}1$ Readys), then it can stop disseminating and discard all Echo messages it might receive that are related to this content. 

\begin{proof}
If $p_i$ Bracha-delivered a message, then $p_i$ has Dolev-delivered Ready messages from $2f{+}1$ distinct processes and at least $2f{+}1$ processes are reliably exchanging the related Ready messages. Thus, all processes will eventually Bracha-deliver the message and the Echo message is not needed. 
\end{proof}

\textbf{Receiving Readys before transmitting Echos [MBD.8].}
If a node $p_i$ has Dolev-delivered the Ready message of its neighbor
$p_j$, it can avoid sending any future Echo message it receives to
$p_j$. Note that this happens as soon as $p_j$'s Ready is received,
since it is transmitted with an empty path and therefore immediately
delivered.

\begin{proof}
If $p_j$ is faulty, whether or not it transmits a message to~$p_j$ will not impact the protocol's guarantees. 
If~$p_j$ is correct, it has Dolev-delivered $\ceil{\frac{N+f{+}1}{2}}$ Echos, or $f{+}1$ Readys,  
before sending its Ready, and each of these Echos or Readys have been forwarded with empty paths. These messages will eventually be Dolev-delivered by all processes. 
\end{proof}

\textbf{Avoiding neighbors that delivered [MBD.9].} If a node~$p_i$ has received $2f{+}1$ Readys (generated by $2f{+}1$ different processes) with empty paths that are related to the same content from a neighbor $p_j$,
then $p_i$ can avoid sending any message related to that content to $p_j$ in the future.

\begin{proof}
If $p_j$ is correct it sent $2f{+}1$ Readys with an empty path to all its neighbors (MD.2). These neighbors will eventually receive these $2f{+}1$ Readys and be able to deliver independently from $p_i$'s message transmissions through $p_j$. Again, if $p_j$ is faulty, it will not make a difference for the protocol's properties whether or not it receives messages from $p_i$.  
\end{proof}

\textbf{Ignore messages whose path is a superpath of a message previously received [MBD.10].}  If a node $p_i$ receives a message $m_1$ (e.g., an Echo from process $p_0$) with path ${path}_1$, for which a previous message $m_0$ with ${path}_0$ was received (and which only differs from $m_1$ by its path) and such that ${path}_0$ is a subpath of ${path}_1$, then $p_i$ can ignore $m_1$. 

\begin{proof}
The path ${path}_1$ does not help $p_i$ identify $2f{+}1$ disjoint
paths towards the source, because ${path}_0$ contains a subset of the
processes that are included in ${path}_1$. Similarly, it is not useful
to forward the received message with a modified path equal to
${path}_1\! +\! [p_i]$, because a subpath such as ${path}_0\! +\!
[p_i]$ or $[\ ]$ has already been transmitted.
\end{proof}

Upon receiving a message, a process checks whether a subpath has been previously received, and if so discard the message, otherwise the received path is saved. Processes represent paths using bit arrays, and store them in a list. The observed performance impact of this method 
does not justify the use of more efficient data structures.  

\subsection{Non-tight cases}

We introduce several modifications that decrease the number of processes that participate in various phases of the Bracha-Dolev protocol combination when the number of processes in the network is larger than $3f{+}1$. 

\textbf{Reduced number of messages in Bracha [MBD.11].}
To implement this method, we assume that processes know the ID of all processes in the system. \newtext{When process $p_i$ BD-broadcasts a message,} the processes with the $\ceil{\frac{N+f{+}1}{2}}\! +\! f$ smallest IDs \newtext{after $p_i$'s (modulo the total number of processes)} generate Echos, while the processes with the $2f\!+\!1\!+\!f$ smallest IDs generate Readys. The other processes simply relay the messages they receive. All processes deliver when they have collected $2f\!+\!1$ Readys. \newtext{Note that the processes that participate actively in a broadcast, i.e., generate new Echo or Ready messages, depend on the ID of the process that initiated the broadcast to distribute the load over all processes.}

\begin{proof}
To realize this, one has to reason starting from the end of Bracha's protocol. 
A process needs $2f{+}1$ Readys to deliver, therefore only $3f{+}1$ processes are required to send a Ready, 
since there are at most $f$ faulty processes in the system, and there are no message losses. To be able to send a Ready, a process needs to receive $\ceil{\frac{N+f{+}1}{2}}$ Echos. Therefore, $\ceil{\frac{N+f{+}1}{2}}\! +\! f$ processes are required to send an Echo. 
Finally, the source needs to transmit its Send message to $\ceil{\frac{N+f{+}1}{2}} + f$ processes. Note that when $N=3f{+}1$, the number of processes that are chosen to participate in each phase is always equal to $3f{+}1$, as indicated by Bracha's protocol.  
\end{proof}

\textbf{Reduced fanout [MBD.12].} \newtext{A process can transmit its newly created Send, Echo or Ready message to only $2f{+}1$ of its neighbors, instead of to all of them.}

\begin{proof} 
\newtext{If a process sends a message to $2f{+}1$ of its neighbors, then at least $f{+}1$ correct of them will eventually receive it. These $f{+}1$ correct nodes will then disseminate the received message with an empty path (cf. MBD.2) to all other nodes in the system. Since every pair of processes are connected through at least $2f{+}1$ disjoint paths, every process in the system will eventually receive the message through at least $f{+}1$ disjoint paths and RC-deliver it.} 
\end{proof}

\subsection{Discussion.} 

Modification MBD.11 drastically decreases the number of messages exchanged. However, it also often increases the protocol's latency. Indeed, the processes selected to generate Echo/Ready messages might be located far from each other in the network. Without this modification all processes send their Echo/Ready messages, which allows processes to collect the required numbers of Echos/Readys faster. \newtext{Note that MBD.12 can be slightly optimized when used in combination with MBD.11: processes can select the $2f{+}1$ neighbors to which they transmit their newly created messages so that they contain the processes that are allowed to create Echo or Ready messages.}

\newtext{In order to Dolev-deliver messages, processes identify $f{+}1$ received disjoint paths based on the brute-force method. However, the algorithm's complexity is slightly reduced by the fact that processes associate the paths they receive with the neighbor that sent them, so that $f{+}1$ paths received from different neighbors have to be found disjoint. In addition, processes use dynamic programming to generate and remember the combinations of disjoint paths they have explored and the number of paths that were used to generate them. A newly received path is then combined with these generated combinations to try to identify $f{+}1$ disjoint paths instead of recomputing all possible combinations.}

\section{Performance evaluation}
\label{sec:perf_eval}

\subsection{Settings}

\newtext{
We implemented our modified Bracha-Dolev protocol combination in {C\nolinebreak[4]\hspace{-.05em}\raisebox{.4ex}{\tiny\bf ++}} using the Salticidae networking library~\cite{SALTICIDAE}. 
We run the experiments on a desktop that runs Ubuntu 20.04 and is equipped with 64~GB of memory and a 12th Gen. Intel(R) Core(TM) i9-12900KF processor. Each node is hosted in a Docker container and communicate using TCP sockets to emulate authenticated channels. We limit the network bandwidth using the netem network utility to 1~Gb/s. We consider both synchronous and asynchronous network settings with the same average message delay to study the effects of network asynchrony. To simulate synchronous networks, each message is delayed by 50~ms. To simulate asynchronous networks, each message is delayed by 50 +/- 50~ms using a normal distribution. To measure the CPU and memory consumption during experiments we rely on the psutil Python library.
In a preliminary version of this work~\cite{bonomi2021practical}, we used the \omnetpp network simulator v.5.6.2~\cite{omnet}. Both versions of the code are available online\footnote{\url{https://github.com/jdecouchant/BRB-on-partially-connected-networks}}.}

We evaluate the impact of modifications MBD.1--12 on BDopt, the state-of-the-art combination of Bracha's algorithm with Dolev's algorithm modified with Bonomi et al.'s improvements~\cite{bonomi2019multi}. 
We vary the number $N$ of processes, the number $f$ of Byzantine processes, 
and the network connectivity $k$, so that $N \ge 3f{+}1$ and $k \ge
2f{+}1$. For each experiment with a $(N, k, f)$ tuple, we generate a regular
random graph~\cite{DBLP:journals/cpc/StegerW99} using the NetworkX Python
library~\cite{hagberg2008exploring}, and report the average of at least 5 runs.

Table~\ref{tab:fields_size} in Appx.~\ref{sec:implemDetails} details the size of the message
fields we use. 
We measure the protocols' latency as the amount of time necessary for all correct processes to deliver the broadcast message.
Note that further messages might still be exchanged after this
point. We consider payloads of 16~B and 1024~B, to represent small and
large messages. In our experiments, only one process broadcasts a message at a time to limit the overall memory consumption and to prevent simultaneous broadcasts from interfering with each other.  

\subsection{Impact of individual modifications}

\begin{sidewaystable*}
\caption{Impact of the modifications with random graphs and synchronous communications. Values for MBD.2--12 are relative to the first configuration where MBD.1 is used. We use '-' when there is no clear tendency whether or not an option should be used depending on parameters $f$ and $k$.} 
\label{tab:impact_mods}
\footnotesize
\begin{center} 
\begin{tabular}{|p{.5cm}|p{.6cm}|p{1.4cm}p{1.7cm}|p{1.4cm}p{1.7cm}||p{1.4cm}p{1.7cm}|p{1.4cm}p{1.7cm}|} 
\cline{3-10} 
 \multicolumn{2}{c}{} & \multicolumn{4}{|c||}{Small payload (16~B)} & \multicolumn{4}{c|}{Large payload (1024~B)} \\
\hline 
\textbf{MBD} & \textbf{Layer} & \textbf{Lat. var. \%} & \textbf{Useful when} & \textbf{\# bits var.} & \textbf{Useful when} & \textbf{Lat. var. \%} & \textbf{Useful when} & \textbf{\# bits var.} & \textbf{Useful when} \\
\hline
1 
& D & $[-57, -19]$ & Always & $[-68, -61]$ & Always & $[-97, -18]$ & Always except high k & $[-98, -97.6]$  &  Always \\
\hline
2  
& BD & $[-7.8, +8.2]$ & Med. and High k & $[-6.5, +0.9]$ & Med. and High k & $[-5.0, +6.0]$ & Always except high k & $[-4.8, +3.4]$ & Always  \\
\hline
3 
& BD &  $[-2.8, +9.6]$ & Low k & $[-4.4, +4.6]$ & Low k & $[-4.6, +3.5]$ &   - & $[-3.2, +5.1]$ & High f \\
\hline
4 
& BD & $[-7.3, +4]$ & - & $[-3.3, +3.9]$ & High f & $[-5.5, +3.7]$ & - & $[-4.6, +2.8]$ & - \\
\hline
5  
& BD & $[-5.9, +6.8]$ & High k & $[-2.7, +3.6]$ & High f & $[-4.3, +5.3]$ & - & $[-4.7, +4.2]$  & - \\
\hline
6 
& B & $[-6.2, +6.6]$ & - & $[-4.1, +5.0]$ & High f & $[-5.1, +3.6]$ & - & $[-4.3, +2.7]$ &  - \\
\hline
7 
& B & $[-20, +3.1]$ & Med. and High f & $[-47, -23]$ & Always & $[-25, +1.1]$ & Always & $[-34, -22]$ &  Always \\
\hline
8  
& B & $[-13, +2.6]$ & Always & $[-16, -4.4]$ & Always & $[-10, +1.4]$ & High f & $[-15, -3.1]$ & Always \\
\hline
9  
& BD & $[-11, +3.6]$ & Always & $[-11, +1.3]$ & Always & $[-5.0, +3.6]$ & Low f & $[-8.6, +2.1]$ &  High f and k\\
\hline
10 
& D & $[-5.1, +3.8]$ & Low k & $[-3.8, +4.9]$ & High k & $[-3.6, +6.2]$ & Med. f & $[-4.8, +3.0]$ & High f and low k \\
\hline
11 
& B & $[-3.6, +35]$ & Never & $[-55, -12]$ & Always & $[-5.7, +25]$ & Med. f and k & $[-32, -8.0]$ & Always except high f and k \\
\hline
12 
& B & $[-2.0, +36]$ & High f & $[-2.4, +13]$ & High f & $[-1.2, +25]$ & - & $[-0.7, +21]$ & - \\
\hline
\end{tabular}
\end{center} 
\end{sidewaystable*}

We evaluated the individual impact of each modification on BDopt's latency and network consumption with a large set of parameters. Table~\ref{tab:impact_mods} summarizes the protocol layer on which a modification is applied, and its average impact on latency and throughput for small and large payload sizes. Given that MBD.1 has such a large impact on network consumption and latency, we report the impact of MBD.2--12 compared to a version of BDOpt that includes MBD.1. 
Tables~\ref{tab:impact_mods_net_sync} and~\ref{tab:impact_mods_lat_sync} in Appendix present a box plot for each modification respectively for network consumption and latency.
We observed that one can decide whether a modification should be used to improve either latency or the network consumption based on the payload size and the network connectivity k. We observed that the value of parameter $f$ does not significantly impact latency or the network consumption.

\begin{figure}
     \centering
     \begin{subfigure}[b]{0.49\textwidth}
       	\centering
	\includegraphics[width=\columnwidth]{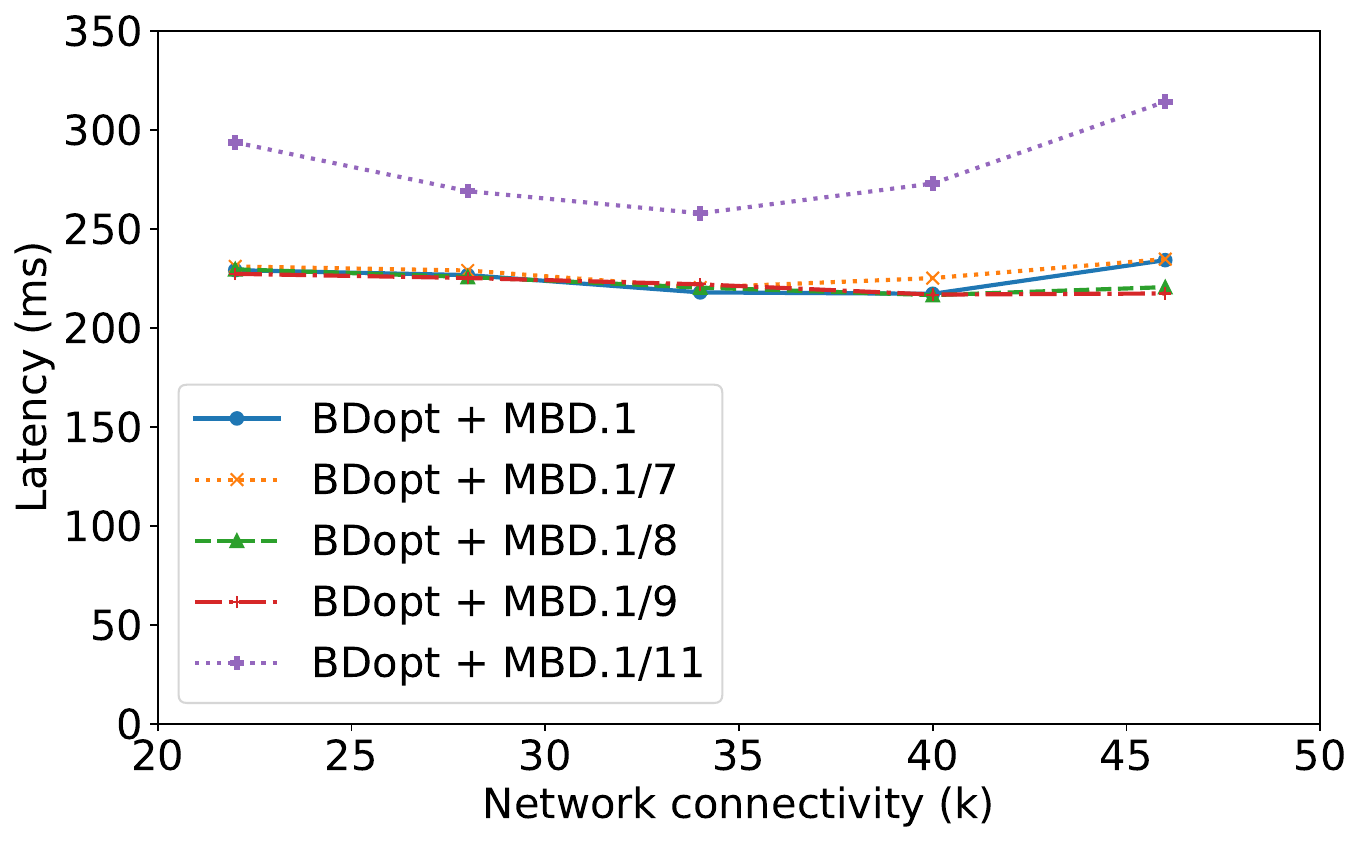}
	\caption{Latency}
	\label{fig:3_lat}
     \end{subfigure}
     \hfill
     \begin{subfigure}[b]{0.49\textwidth}
\includegraphics[width=\columnwidth]{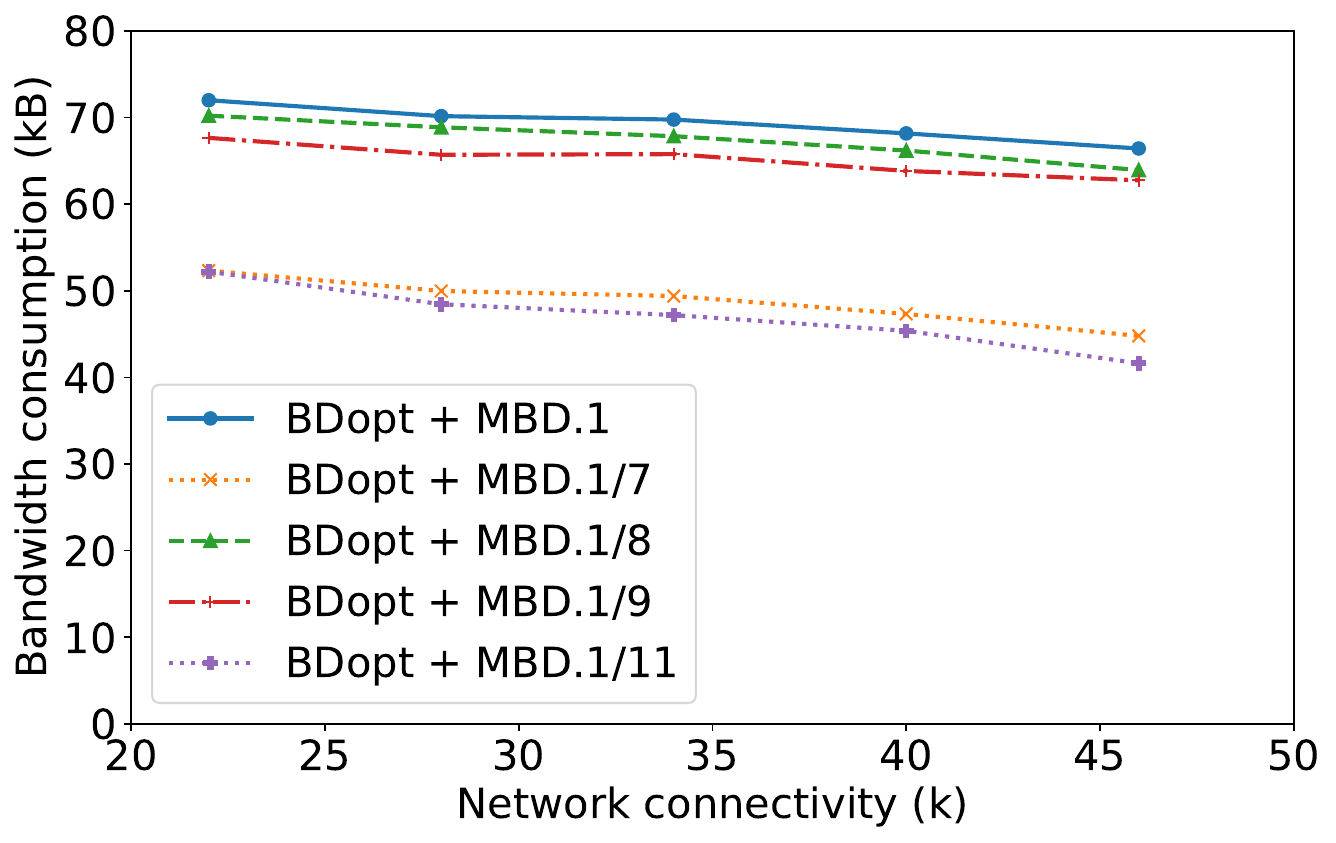}
	\caption{Network consumption}
	\label{fig:8_lat}
     \end{subfigure}
        \caption{Modifications MBD.1, 7, 8, 9, 11 - $N{=}50, f{=}9$, 1024~B payload.}
\end{figure}

We illustrate the impacts of some selected modifications when $N=50$, $f=9$ and using a 1024~B payload in Fig.~\ref{fig:3_lat} and~\ref{fig:8_lat} on latency and network consumption, respectively.
At this scale, the latency obtained thanks to the use of MBD.1 appears to be independent of the network connectivity (remember that MBD.1 already decreases latency by up to 97\%).
First, Fig.~\ref{fig:3_lat} shows that MBD.7, 8, 9 do not modify the latency to a large extent compared to the use of MBD.1 alone, which decreased it down to around 240~ms, while MBD.11 increases the latency to up to 280-300ms. We recall that MBD.11 limits the number of processes that generate Echo/Ready messages while the others only relay messages.

Fig.~\ref{fig:8_lat} shows the impact of MBD.7, 8, 9 and 11 on network consumption. After MBD.1, which reduces the network consumption to around 70~KB, MBD.7 and MBD.11 are the two options that decrease the network consumption the most, both to around 50~KB. 
From Table~\ref{tab:impact_mods}, one can also observe that larger payloads lead to larger performance improvements.

\subsection{CPU and memory consumption}

We measured the total CPU and memory consumption  
with 10, 30, and 50 nodes and a 16~B payload, for a wide range of possible $f$ and network connectivity values.  
In the synchronous settings, processes use at most 47~MB, 75~MB and 618~MB when $N$ is respectively equal to 10, 30 and 50. 
The memory usage increases significantly with the size of the system and comes from the storage of the received transmission paths that need to be processed to authenticate a message. 
This is expected since the number of paths received for each Dolev message that a reliable broadcast involves (without any modifications) increases exponentially with the number of processes.

\subsection{Latency vs. Network Consumption}

\begin{figure}
     \centering
     \begin{subfigure}[b]{0.49\textwidth}
           \centering
	\includegraphics[width=\columnwidth]{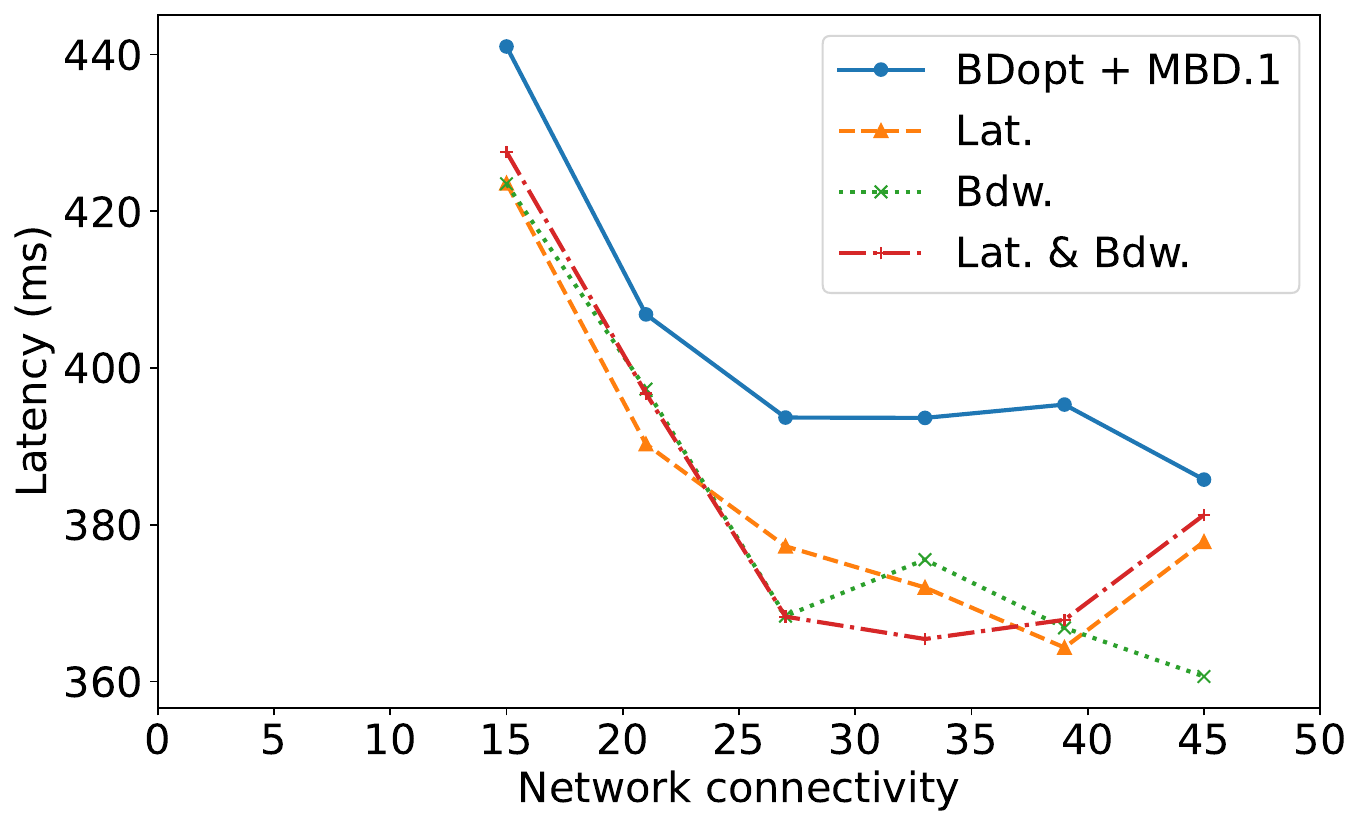}
	\caption{Latency}
	\label{fig:4_lat_raw_small}
     \end{subfigure}
     \hfill
     \begin{subfigure}[b]{0.49\textwidth}
	\centering
	\includegraphics[width=\columnwidth]{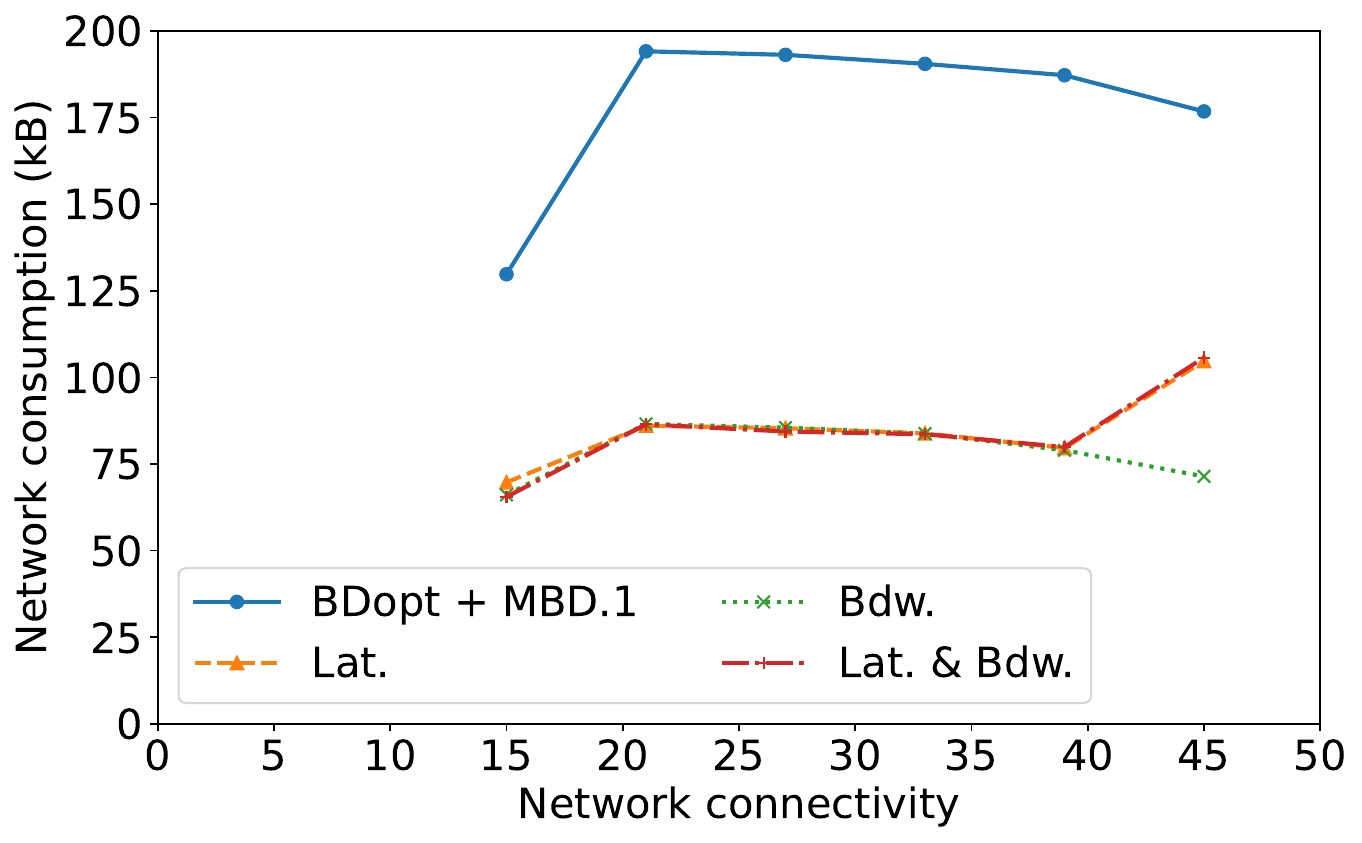}
	\caption{Bandwidth consumption}
	\label{fig:4_bdw_raw_small}
     \end{subfigure}
        \caption{Configuration - $(N,f)\!=\!(50,10)$, 1024~B payload.}
\end{figure}

Based on the observed impact of our modifications, we compare the latency and network consumption improvements of three combinations of modifications to BDopt with MBD.1 that respectively contain:
(i) \textit{lat.}: only the modifications that decrease latency; 
(ii) \textit{bdw.}: only the modifications that decrease bandwidth consumption; and
(iii) \textit{lat. \& bdw.}: only the modifications that decrease both latency and bandwidth consumption. 
Figs.~\ref{fig:4_lat_raw_small} and~\ref{fig:4_bdw_raw_small} respectively show the latency decrease and the network consumption of these configurations with $(N,f)\!=\!(50,10)$ and a 1024~B payload. All three complex configurations, i.e., \textit{lat.},  \textit{bdw.} and \textit{lat. \& bdw.} always decrease latency, e.g., from around 400~KB to 370~KB when $k=30$. 
All three configurations decrease the network consumption in a very similar way, e.g., from around 190~KB to around 90~KB when $k=30$, which demonstrates that options MBD.2--12 significantly decrease the network consumption after MBD.1.  

\newtext{
These results show that there is no simple combination of the MBD.1--12 modifications that optimizes latency, while optimizing for network consumption is more straightforward. However, Figs.~\ref{fig:scal_bits_small} and~\ref{fig:scal_lat_small} show that the \textit{lat.} and \textit{bdw.} configurations behave as expected with $N=30$. We evaluated every possible combination of optimizations and could not identify a combination of modifications that always performed better than the others. This suggests that our modifications sometimes have complex interactions. A user aiming at obtaining the best possible latency or network consumption under particular settings might therefore want to benchmark possibly interesting combinations.
}

\subsection{Evolution with the number of processes}

\begin{figure}
     \centering
     \begin{subfigure}[b]{0.49\textwidth}
            \centering
            \includegraphics[width=\columnwidth]{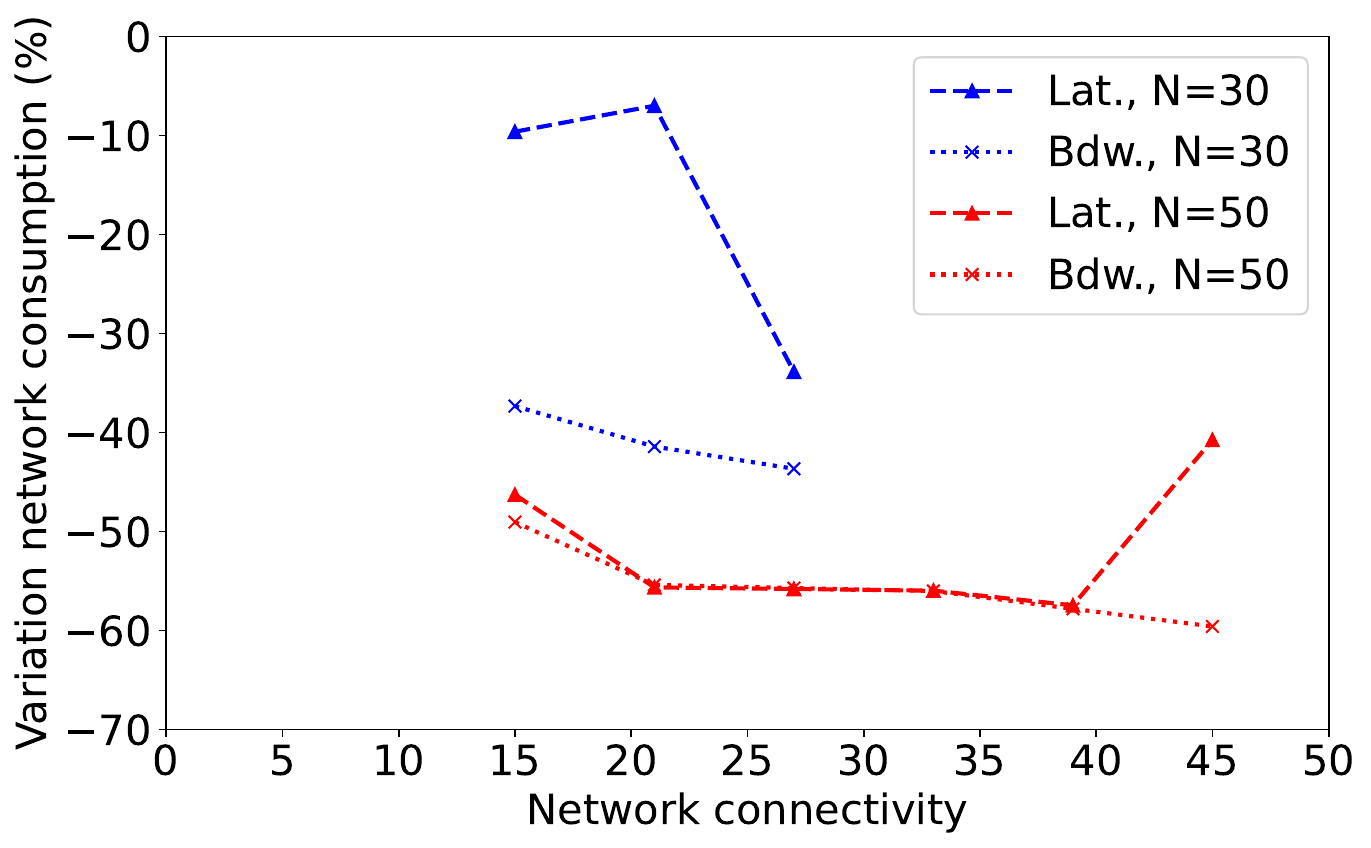}
    	\caption{Bandwidth consumption improvement}
    	\label{fig:scal_bits_small}
     \end{subfigure}
     \hfill
     \begin{subfigure}[b]{0.49\textwidth}
         \centering
    	\includegraphics[width=\columnwidth]{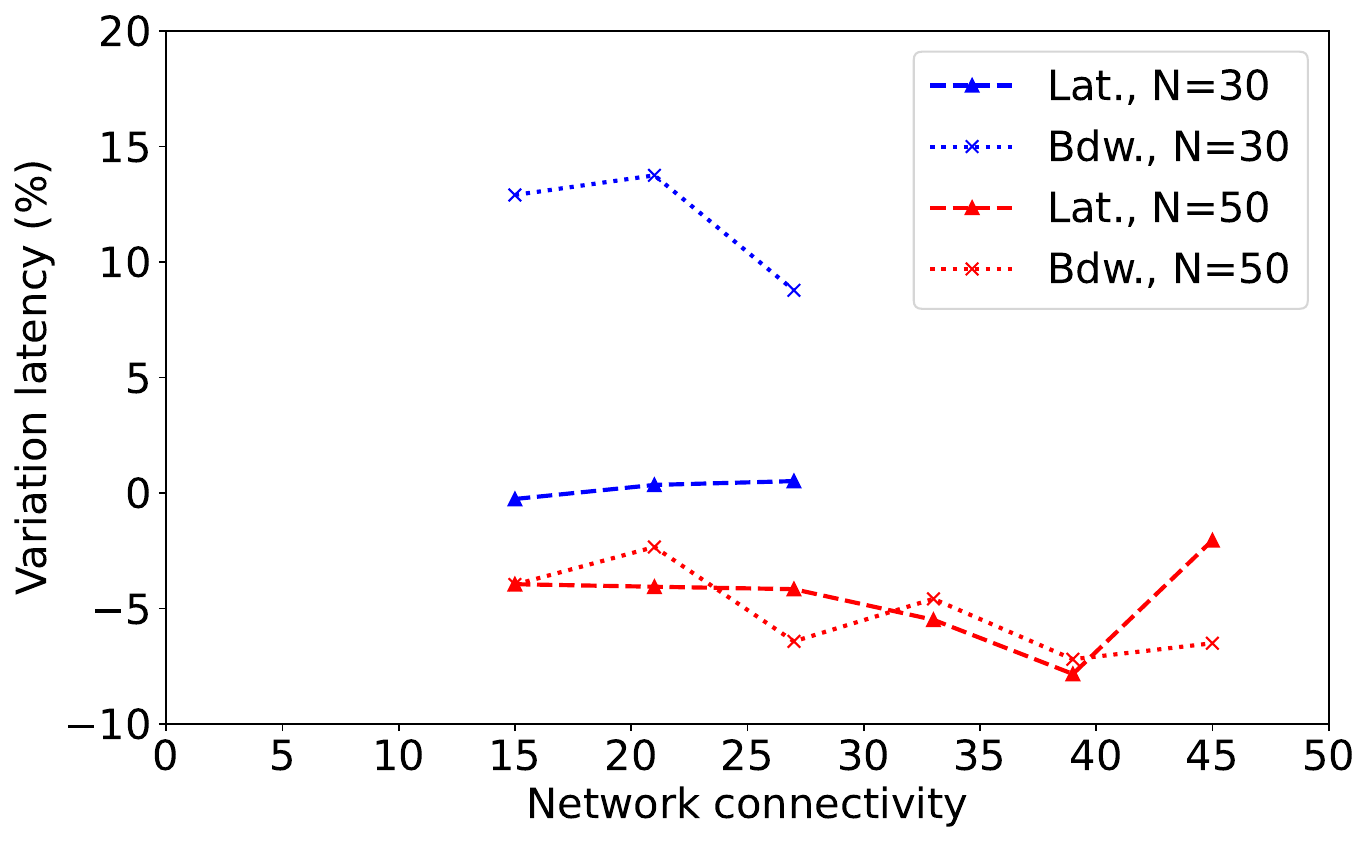}
    	\caption{Latency improvement}
    	\label{fig:scal_lat_small}
     \end{subfigure}
        \caption{BDopt - 1024~B payload.}
\end{figure}

In this experiment, we considered graphs of 30 or 50 processes. 
We chose the average possible $f$ value for each $N$ value, since we have seen that $f$ does not significantly modify the network consumption or protocol latency. Using larger $N$ values is challenging because of the exponential number of messages, memory and computing time that processes require to execute the Bracha-Dolev protocol combination.  Figs.~\ref{fig:scal_bits_small} and~\ref{fig:scal_lat_small} respectively show the bandwidth and latency improvement of the \textit{lat.} and \textit{bdw.} configurations over BDopt with MBD.1 activated with a 1024~B payload. We omit configuration \textit{lat. \& bdw.} in those figures for clarity.
As one can see, configurations \textit{bdw.} and \textit{lat.} with $N=50$ further improve the bandwidth consumption respectively by around -55\%. With $N=30$, the \textit{bdw.} configuration improves the bandwidth consumption by around -40\% wile the \textit{lat.} configuration reduces it by -10\% to -30\%. 
Configuration \textit{lat.} and \textit{bdw.} improve latency when $N=50$ by around -5\%, but fail to do so when $N=30$, where only the \textit{lat.} configuration maintain the same latency as with MBD.1 only. 

\subsection{Asynchronous networks}

In addition to the above experiments, we also performed experiments with 50 +/- 50~ms transmission delays, which were computed per message using a normal distribution. With these settings, messages were frequently reordered during their transmission.
Tables~\ref{tab:impact_mods_net_async} and~\ref{tab:impact_mods_lat_async} in Appendix detail the impact of each modification under our settings.
Modifications MBD.1--12 also perform well in asynchronous networks, but seem to have a slightly lower impact than with synchronous networks (e.g., MBD.11 decreases network consumption by 18\% instead of 24\%) and exhibit higher variation. 

\subsection{Comparison with the results obtained with the \omnetpp network simulator}

\newtext{Compared to our previous performance evaluation that used a network simulator~\cite{bonomi2021practical} instead of real code deployment, we have found that modifications MBD.1-12 decrease latency by a smaller extent. 
A possible explanation is that local computations made upon the receipt of each message are no longer abstracted, but concretely executed.
However, our modifications do improve latency in cases where messages are delayed because the network is saturated by decreasing the amount of transmitted data, and allow nodes to transition to the next phase slightly faster. Compared to our evaluation based on a network simulator, we observed that in a real deployment combining the modifications to optimise either latency or network consumption is non trivial, as the impact of a combination of individual modifications is not always the sum of their individual impact, and requires careful benchmarking.}

\section{Conclusion} 
\label{sec:conclusion}

In this paper, we discussed the practicality of Byzantine reliable broadcast problem on unknown partially connected topologies. From a theory perspective, this problem can be solved by combining two seminal protocols, Dolev's and Bracha's algorithms. We first explained how recent improvements of Dolev's algorithms can be used in this protocol combination. We then described a total of 12 new modifications one can apply to this protocol combination to optimize latency and/or bandwidth consumption. We benchmarked the impact of each modification using an actual C++ based implementation of our proposal, demonstrating its practical feasibility and its adaptability for specific latency and/or bandwidth requirements. Our reference implementation is publicly available to the community, in the hope additional applications can benefit from it, and that further improvements can be proposed by the research community.
Future works include to also consider the local fault model used for example in the CPA line of work~\cite{koo2004broadcast,pelc2005broadcasting,litsas2013graph}.

\printbibliography

\appendix

\section{Summary of modifications and implementation details}
\label{sec:implemDetails}

Table~\ref{tab:summary_modifications} in recalls MD.1--5, the state-of-the-art modifications of Dolev's protocol~\cite{bonomi2019multi}, and the novel modifications MBD.1-12 we present in this work. We indicate for each modification the layer(s) it applies to. We use D to refer to Dolev's RC protocol; B to refer to Bracha's BRB protocol; BD to refer to the Bracha-Dolev protocol combination; and BDopt to refer to the version of BD that is optimized with modifications MD.1--5.

\begin{table}[htbp]
\footnotesize
\begin{center} 
\caption{Summary of modifications.}
\label{tab:summary_modifications}
\begin{tabular}{|r|p{4.8cm}|p{1.2cm}|} 
\hline 
\textbf{MD} & \textbf{Description} & \textbf{Layer}\\
\hline
MD.1 & Directly deliver message received from source. & D\\
MD.2 & Relay empty path upon delivery. & D \\
MD.3 & Do not send to neighbors that delivered. & D \\ 
MD.4 & Do not relay or analyze paths that contain the ID of a process that delivered. & D \\
MD.5 & Stop after delivering. & D \\
\hline
\hline
MBD.1 & Use local ID for payload. & D\\
MBD.2 & Single-hop Send messages. & BD \\ 
MBD.3 & Echo-to-echo transition merge. & BD \\ 
MBD.4 & Echo-to-ready transition merge. & BD \\ 
MBD.5 & Optimized message format. & BD \\ 
MBD.6 & Ignore echos after ready. & B \\
MBD.7 & Ignore echos after delivering. & B \\
MBD.8 & Do not send echo to ready neighbors. & B \\
MBD.9 & Do not send to neighbors that delivered. & BD \\ 
MBD.10 & Ignore messages whose path is a superpath of a known message. & D \\
MBD.11 & Overprovisioning in Bracha. & B \\
MBD.12 & Send message to $2f{+}1$ neighbors. & B \\
\hline
\end{tabular}
\end{center} 
\end{table} 

Table~\ref{tab:fields_size} details the size of the message
fields we use in our \cpp implementation. 

\begin{table}[h]
\footnotesize
\begin{center} 
\caption{\newtext{Description and size of the message fields.}}
\label{tab:fields_size}
\begin{tabular}{|p{2cm}|p{4cm}|p{1.3cm}|} 
\hline 
\textbf{Msg field} & \textbf{Description} & \textbf{Size}\\
\hline
$mtype$ & Message type & 1~B \\
$s$ & ID of the source process & 4~B \\
$bid$ & Message ID & 4~B \\
$localPayloadID$ & Local ID for payload & 4~B\\
$payloadSize$ & Payload data size in Bytes & 4~B \\
$payload$ & Payload data & 1024~B \\
${erId}_1$ & Echo/Ready sender ID & 4~B \\
${erId}_2$ & Embedded Echo/Ready sender ID (only Echo\_Echo and Ready\_Echo msgs) & 4~B \\
$pathLen$ & Path length (number of processes) & 2~B \\
$path$ & List of process IDs & 4~B per ID \\
\hline
\end{tabular}
\end{center} 
\end{table} 

\section{Detailed impact of individual modifications with synchronous and asynchronous networks}

Figures~\ref{tab:impact_mods_net_sync} and~\ref{tab:impact_mods_net_async} respectively present the impact on network consumption (in \%) of each individual modification of the base Bracha-Dolev protocol combination with 1~KiB payloads, 50 processes and random graphs over all our experiments. The box plots report the 95\% interval, the quartiles and the median impact, which are indicated on the side of the figure.  In bose cases, the five most important modifications for network consumption are, in decreasing order of importance, MBD.1, MBD.7, MBD.11, MBD.8 and MBD.9. Network asynchrony increases the spread of each modification and reduces the impact of each modification. 

Figures~\ref{tab:impact_mods_lat_sync} and~\ref{tab:impact_mods_lat_async} respectively present the impact on latency (in \%) of each individual modification of the base Bracha-Dolev protocol combination, with the same experimental settings. In bose cases, the five most important modifications for latency are, in decreasing order of importance, MBD.1, MBD.7, MBD.8, MBD.9 and MBD.2. Contrary to the observed impact on network consumption, the use of modifications in presence of network asynchrony seems to increase the impact and reduce the spread of each modification. 

\begin{figure*}[h]
\centering
\resizebox{\textwidth}{!}{%
\begin{tikzpicture}
\pgfplotstablegetrowsof{\datatablenetsync}
\pgfmathtruncatemacro{\rownumber}{\pgfplotsretval-1}
\begin{axis}[boxplot/draw direction=x,yticklabels={[-98.8 -98.1 -$\mathbf{97.8}$ -97.6 -96.8] MBD.1, [-17 -4.8 -$\mathbf{0.8}$ 3.4 16] MBD.2, [-16 -3.2 $\mathbf{0.9}$ 5.1 18] MBD.3, [-16 -4.6 -$\mathbf{0.3}$ 2.8 14] MBD.4, [-18 -4.7 -$\mathbf{0.3}$ 4.2 18] MBD.5, [-15 -4.3 -$\mathbf{0.3}$ 2.7 13] MBD.6, [-51 -34 -$\mathbf{29}$ -22 -6] MBD.7, [-32 -15 -$\mathbf{8.8}$ -3.1 14] MBD.8, [-25 -8.6 -$\mathbf{4.3}$ 2.1 18] MBD.9, [-17 -4.8 -$\mathbf{0.3}$ 3.0 15] MBD.10, [-69 -32 -$\mathbf{25}$ -7.9 29] MBD.11, [-33 -0.7 $\mathbf{7.0}$ 21 53] MBD.12},ytick={1,...,\the\numexpr\rownumber+1},height=10cm]
\typeout{\rownumber}a
\pgfplotsinvokeforeach{0,...,\rownumber}{
 \pgfplotstablegetelem{#1}{low}\of\datatablenetsync
 \edef\mylow{\pgfplotsretval}
 \pgfplotstablegetelem{#1}{ql}\of\datatablenetsync
 \edef\myql{\pgfplotsretval}
 \pgfplotstablegetelem{#1}{med}\of\datatablenetsync
 \edef\mymed{\pgfplotsretval}
 \pgfplotstablegetelem{#1}{qh}\of\datatablenetsync
 \edef\myqh{\pgfplotsretval}
 \pgfplotstablegetelem{#1}{high}\of\datatablenetsync
 \edef\myhigh{\pgfplotsretval}
 \typeout{\mylow,\myql,\mymed,\myqh,\myhigh}
 \edef\temp{\noexpand\addplot+[
    boxplot prepared={
     lower whisker=\mylow,
     upper whisker=\myhigh,
     lower quartile=\myql,
     upper quartile=\myqh,
     median=\mymed
  }
  ]coordinates {};}
 \temp
}
\end{axis}
\end{tikzpicture}
}%
\caption{Impact (in \%) of a single modification with random graphs and \textbf{synchronous} communications on network consumption (payload 1~KiB).} 
\label{tab:impact_mods_net_sync}
\end{figure*}

\begin{figure*}[h]
\centering
\resizebox{\textwidth}{!}{%
\begin{tikzpicture}
\pgfplotstablegetrowsof{\datatablenetasync}
\pgfmathtruncatemacro{\rownumber}{\pgfplotsretval-1}
\begin{axis}[boxplot/draw direction=x,yticklabels={[-100 -97.5 -$\mathbf{96.8}$ -95.7 -93.0] MBD.1, [-30 -8.6 -$\mathbf{1.9}$ 5.4 26] MBD.2, [-37 -11 $\mathbf{0.5}$ 6.6 33] MBD.3, [-32 -7.8 $\mathbf{0.4}$ 8.3 32] MBD.4, [-20 -3.5 $\mathbf{1.2}$ 7.4 24] MBD.5, [-23 -6.5 $\mathbf{0.1}$ 4.5 21] MBD.6, [-55 -30 -$\mathbf{24}$ -13 13] MBD.7, [-34 -13 -$\mathbf{6.7}$ 0.4 21] MBD.8, [-34 -9.2 -$\mathbf{4.3}$ 7.0 31] MBD.9, [-34 -10 -$\mathbf{0.7}$ 5.6 29] MBD.10, [-86 -34 -$\mathbf{18}$ 0.5 52] MBD.11, [-36 -4.1 $\mathbf{5.1}$ 17 49] MBD.12},ytick={1,...,\the\numexpr\rownumber+1},height=10cm
                ]
\typeout{\rownumber}
\pgfplotsinvokeforeach{0,...,\rownumber}{
 \pgfplotstablegetelem{#1}{low}\of\datatablenetasync
 \edef\mylow{\pgfplotsretval}
 \pgfplotstablegetelem{#1}{ql}\of\datatablenetasync
 \edef\myql{\pgfplotsretval}
 \pgfplotstablegetelem{#1}{med}\of\datatablenetasync
 \edef\mymed{\pgfplotsretval}
 \pgfplotstablegetelem{#1}{qh}\of\datatablenetasync
 \edef\myqh{\pgfplotsretval}
 \pgfplotstablegetelem{#1}{high}\of\datatablenetasync
 \edef\myhigh{\pgfplotsretval}
 \typeout{\mylow,\myql,\mymed,\myqh,\myhigh}
 \edef\temp{\noexpand\addplot+[
    boxplot prepared={
     lower whisker=\mylow,
     upper whisker=\myhigh,
     lower quartile=\myql,
     upper quartile=\myqh,
     median=\mymed
  }
  ]coordinates {};}
 \temp
}
\end{axis}
\end{tikzpicture}
}%
\caption{Impact (in \%) of a single modification with random graphs and \textbf{asynchronous} communications on network consumption (payload 1~KiB).} 
\label{tab:impact_mods_net_async}
\end{figure*}

\begin{figure*}[h]
\centering
\resizebox{\textwidth}{!}{%
\begin{tikzpicture}
\pgfplotstablegetrowsof{\datatablelatsync}
\pgfmathtruncatemacro{\rownumber}{\pgfplotsretval-1}
\begin{axis}[boxplot/draw direction=x,yticklabels={[-99.8 -98.1 -$\mathbf{97.8}$ -97.6 -96.8] MBD.1, [-17 -4.8 -$\mathbf{0.8}$ 3.4 16] MBD.2, [-16 -3.2 $\mathbf{0.9}$ 5.1 18] MBD.3, [-16 -4.6 -$\mathbf{0.3}$ 2.8 14] MBD.4, [-18 -4.7 -$\mathbf{0.3}$ 4.2 18] MBD.5, [-15 -4.3 -$\mathbf{0.3}$ 2.7 13] MBD.6, [-50 -34 -$\mathbf{29}$ -22 -5.6] MBD.7, [-32 -15 -$\mathbf{8.8}$ -3.1 14] MBD.8, [-25 -8.6 -$\mathbf{4.3}$ 2.1 18] MBD.9, [-17 -4.8 -$\mathbf{0.3}$ 3.0 15] MBD.10, [-69 -32 -$\mathbf{25}$ -7.9 29] MBD.11, [-33 -0.7 $\mathbf{7.0}$ 21 53] MBD.12},ytick={1,...,\the\numexpr\rownumber+1},height=10cm
                ]
\typeout{\rownumber}
\pgfplotsinvokeforeach{0,...,\rownumber}{
 \pgfplotstablegetelem{#1}{low}\of\datatablelatsync
 \edef\mylow{\pgfplotsretval}
 \pgfplotstablegetelem{#1}{ql}\of\datatablelatsync
 \edef\myql{\pgfplotsretval}
 \pgfplotstablegetelem{#1}{med}\of\datatablelatsync
 \edef\mymed{\pgfplotsretval}
 \pgfplotstablegetelem{#1}{qh}\of\datatablelatsync
 \edef\myqh{\pgfplotsretval}
 \pgfplotstablegetelem{#1}{high}\of\datatablelatsync
 \edef\myhigh{\pgfplotsretval}
 \typeout{\mylow,\myql,\mymed,\myqh,\myhigh}
 \edef\temp{\noexpand\addplot+[
    boxplot prepared={
     lower whisker=\mylow,
     upper whisker=\myhigh,
     lower quartile=\myql,
     upper quartile=\myqh,
     median=\mymed
  }
  ]coordinates {};}
 \temp
}
\end{axis}
\end{tikzpicture}
}%
\caption{Impact (in \%) of a single modification with random graphs and \textbf{synchronous} communications on broadcast latency (payload 1~KiB).} 
\label{tab:impact_mods_lat_sync}
\end{figure*}

\begin{figure*}[h]
\centering
\resizebox{\textwidth}{!}{%
\begin{tikzpicture}
\pgfplotstablegetrowsof{\datatablelatasync}
\pgfmathtruncatemacro{\rownumber}{\pgfplotsretval-1}
\begin{axis}[boxplot/draw direction=x,yticklabels={[-175 -94  -$\mathbf{83}$ -39 42] MBD.1, [-29 -8.6 -$\mathbf{2.1}$ 5.1 26] MBD.2, [-26 -6.4 $\mathbf{1.2}$ 6.6 26] MBD.3, [-24 -5.5 -$\mathbf{0.3}$ 6.7 25] MBD.4, [-27 -6.2 $\mathbf{0.6}$ 7.9 29] MBD.5, [-22 -5.1 -$\mathbf{0.6}$ 6.1 23] MBD.6, [-53 -25 -$\mathbf{17}$ -5.8 22] MBD.7, [-33 -13 -$\mathbf{4.4}$ -0.1 19] MBD.8, [-29 -8.7 -$\mathbf{2.6}$ 5.1 26] MBD.9, [-23 -5.3 -$\mathbf{0.1}$ 6.3 24] MBD.10, [-43 -11 $\mathbf{0.2}$ 10 42] MBD.11, [-37 -2.2 $\mathbf{6.5}$ 21 57] MBD.12},ytick={1,...,\the\numexpr\rownumber+1},height=10cm
                ]
\typeout{\rownumber}
\pgfplotsinvokeforeach{0,...,\rownumber}{
 \pgfplotstablegetelem{#1}{low}\of\datatablelatasync
 \edef\mylow{\pgfplotsretval}
 \pgfplotstablegetelem{#1}{ql}\of\datatablelatasync
 \edef\myql{\pgfplotsretval}
 \pgfplotstablegetelem{#1}{med}\of\datatablelatasync
 \edef\mymed{\pgfplotsretval}
 \pgfplotstablegetelem{#1}{qh}\of\datatablelatasync
 \edef\myqh{\pgfplotsretval}
 \pgfplotstablegetelem{#1}{high}\of\datatablelatasync
 \edef\myhigh{\pgfplotsretval}
 \typeout{\mylow,\myql,\mymed,\myqh,\myhigh}
 \edef\temp{\noexpand\addplot+[
    boxplot prepared={
     lower whisker=\mylow,
     upper whisker=\myhigh,
     lower quartile=\myql,
     upper quartile=\myqh,
     median=\mymed
  }
  ]coordinates {};}
 \temp
}
\end{axis}
\end{tikzpicture}
}%
\caption{Impact (in \%) of a single modification with random graphs and \textbf{asynchronous} communications on the network consumption (payload 1~KiB).} 
\label{tab:impact_mods_lat_async} 
\end{figure*}

\end{document}